\documentclass{emulateapj}
\usepackage{apjfonts,rotating}

\def\asca{{\it ASCA\/}}

\def\chandra{{\it Chandra\/}}

\def\hst{{\it {\it HST}\/}}

\def\herschel{{\it Herschel\/}}

\def\xray{\hbox{X-ray}}

\def\etal{{et\,al.}}

\def\ltsima{$\; \buildrel < \over \sim \;$}
\def\simlt{\lower.5ex\hbox{\ltsima}}
\def\gtsima{$\; \buildrel > \over \sim \;$}
\def\simgt{\lower.5ex\hbox{\gtsima}}
\def\kms{\ifmmode{~{\rm km~s^{-1}}}\else{~km s$^{-1}$}\fi}
\def\lsim{\lower0.3em\hbox{$\,\buildrel <\over\sim\,$}}
\def\gsim{\lower0.3em\hbox{$\,\buildrel >\over\sim\,$}}

\def\h2{H$_2$}
\def\flux{ergs~cm$^{-2}$~s$^{-1}$}
\def\lum{ergs~s$^{-1}$}

\def\arcmin{\mbox{$^\prime$}}
\def\sfr{$M_{\odot}$ yr$^{-1}$}

\def\aap{A\&A}
\def\apj{ApJ}
\def\apjl{ApJL}
\def\apjs{ApJS}
\def\aj{AJ}
\def\mnras{MNRAS}
\def\araa{ARA\&A}

\def\nat{Nature}

\begin{document}

\shortauthors{LEHMER ET AL.}
\shorttitle{Concurrent SMBH and Galaxy Growth in $z = 2.23$ HAEs}

%
\title{Concurrent Supermassive Black Hole and Galaxy Growth: Linking Environment and Nuclear Activity in \lowercase{$z$} = 2.23 H$\alpha$ Emitters}
%

\author{
B.~D.~Lehmer,\altaffilmark{1,2}
A.~B.~Lucy,\altaffilmark{2,3}
D.~M.~Alexander,\altaffilmark{4}
P.~N.~Best,\altaffilmark{5}
J.~E.~Geach,\altaffilmark{6}
C.~M.~Harrison,\altaffilmark{4}
A.~E.~Hornschemeier,\altaffilmark{1,2}
Y.~Matsuda,\altaffilmark{7}
J.~R.~Mullaney,\altaffilmark{4}
Ian~Smail,\altaffilmark{8}
D.~Sobral,\altaffilmark{9}
\& 
A.~M.~Swinbank\altaffilmark{4}
}

\altaffiltext{1}{The Johns Hopkins University, Homewood Campus, Baltimore, MD 21218, USA}
\altaffiltext{2}{NASA Goddard Space Flight Centre, Code 662, Greenbelt, MD 20771, USA} 
\altaffiltext{3}{Homer L. Dodge Department of Physics and Astronomy, The University of Oklahoma, 440 W. Brooks St., Norman, OK 73019}
\altaffiltext{4}{Department of Physics, Durham University, South Road, Durham, DH1 3LE, UK}
\altaffiltext{5}{SUPA, Institute for Astronomy, Royal Observatory of Edinburgh, Blackford Hill, Edinburgh EH9 3HJ, UK}
\altaffiltext{6}{Department of Physics, McGill University, 3600 rue University, Montr\'{e}al, Qu\'{e}bec, H3A 2T8, Canada}
\altaffiltext{7}{Chile Observatory, National Astronomical Observatory of Japan, Tokyo 181-8588, Japan}
\altaffiltext{8}{Institute for Computational Cosmology, Durham University, South Road, Durham, DH1 3LE, UK}
\altaffiltext{9}{Leiden Observatory, Leiden University, P.O. Box 9513, NL-2300 RA Leiden, The Netherlands}

%
\begin{abstract}
%

We present results from a $\approx$100~ks \chandra\ observation of the
2QZ~Cluster~1004+00 structure at $z = 2.23$ (hereafter, 2QZ~Clus).  2QZ~Clus
was originally identified as an overdensity of four optically-selected QSOs at
$z = 2.23$ within a $15 \times 15$~arcmin$^2$ region.  Narrow-band imaging in
the near-IR (within the $K$ band) revealed that the structure contains an
additional overdensity of 22 $z = 2.23$ H$\alpha$-emitting galaxies (HAEs),
resulting in 23 unique $z = 2.23$ HAEs/QSOs (22 within the \chandra\ field of
view).  Our \chandra\ observations reveal that 3 HAEs in addition to the 4 QSOs
harbor powerfully accreting supermassive black holes (SMBHs), with
\hbox{2--10~keV} luminosities of \hbox{$\approx$(8--60)~$\times 10^{43}$~\lum}
and \xray\ spectral slopes consistent with unobscured AGN.  Using a large
comparison sample of 210 $z = 2.23$ HAEs in the \chandra-COSMOS field
(\hbox{C-COSMOS}), we find suggestive evidence that the AGN fraction increases
with local HAE galaxy density.  The 2QZ~Clus HAEs reside in a moderately
overdense environment (a factor of $\approx$2 times over the field), and after
excluding optically-selected QSOs, we find the AGN fraction is a factor of
\hbox{$\approx$$3.5^{+3.8}_{-2.2}$} times higher than C-COSMOS HAEs in similar
environments.  Using stacking analyses of the \chandra\ data and \herschel\
SPIRE observations at 250$\mu$m, we respectively estimate mean SMBH accretion
rates ($\dot{M}_{\rm BH}$) and star-formation rates (SFRs) for the 2QZ~Clus and
C-COSMOS samples.  We find that the mean 2QZ~Clus HAE stacked \xray\ luminosity
is QSO-like (\hbox{$L_{\rm 2-10~keV} \approx$~[6--10]~$\times 10^{43}$~\lum}),
and the implied $\dot{M}_{\rm BH}$/SFR~$\approx$~(1.6--3.2)~$\times 10^{-3}$ is
broadly consistent with the local $M_{\rm BH}$/$M_\star$ relation and $z
\approx 2$ \xray\ selected AGN.  In contrast, the C-COSMOS HAEs are on average
an order of magnitude less \xray\ luminous and have $\dot{M}_{\rm
BH}$/SFR~$\approx$~(0.2--0.4)~$\times 10^{-3}$, somewhat lower than the local
$M_{\rm BH}$/$M_\star$ relation, but comparable to that found for $z
\approx$~1--2 star-forming galaxies with similar mean \xray\ luminosities.  We
estimate that a periodic QSO phase with duty cycle $\approx$2--8\% would be
sufficient to bring star-forming galaxies onto the local $M_{\rm BH}$/$M_\star$
relation.  This duty cycle is broadly consistent with the observed C-COSMOS HAE
AGN fraction ($\approx$0.4--2.3\%) for powerful AGN with $L_{\rm X} \simgt
10^{44}$~\lum.  Future observations of 2QZ~Clus will be needed to identify key
factors responsible for driving the mutual growth of the SMBHs and galaxies.

%
\end{abstract}
%


%
\section{Introduction}
%

%
%
\begin{figure*}
\figurenum{1}
\centerline{
\includegraphics[width=12cm]{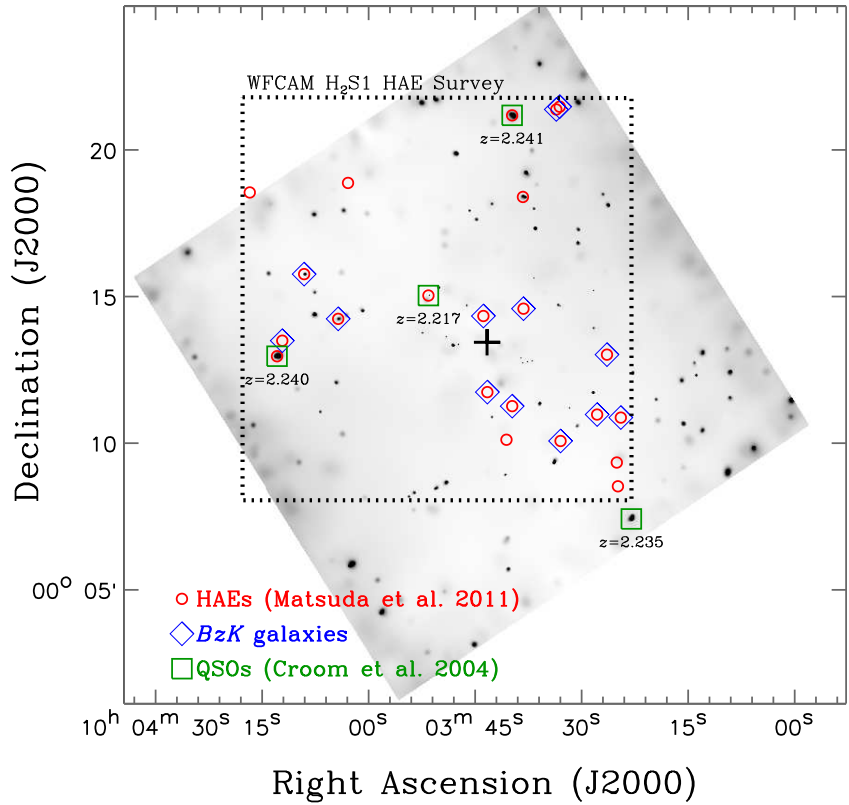}
}
\caption{
Evidence for wide-spread AGN activity for $z = 2.23$ HAEs in the 2QZ~Clus
structure.  We show the adaptively-smoothed \hbox{0.5--7~keV} \chandra\ image
of the 2QZ~Clus field.  For illustrative purposes, the image was binned by a
factor of 4 (in both RA and DEC) and has been exposure-corrected by dividing
the smoothed image by a smoothed exposure map (see $\S$2.2).  The aim point of
the image has been indicated with a cross.  The \hbox{$z =$~2.216--2.248} QSOs
that were used to select the 2QZ~Clus field have been highlighted with green
squares (Croom \etal\ 2001, 2004).  The WFCAM HAE survey region ({\it dashed
square\/}) and $z = 2.23$ source candidates ({\it red circles\/}) have
been highlighted (see Matsuda \etal\ 2011).  $BzK$ candidates have been
highighted with blue diamonds; these sources have colors consistent with
actively star-forming galaxies at \hbox{$z \approx$~1.4--2.5}. 
}
\end{figure*}

Successful theoretical models characterizing the formation and growth history
of galaxies and supermassive black holes (SMBHs) generally require that
feedback from active galactic nuclei (AGN) play a crucial role in regulating
the growth of galaxy bulges (e.g., De~Lucia \etal\ 2005; Bower \etal\ 2006,
2008; Croton \etal\ 2006; Fanidakis \etal\ 2012), leading to the local
$M$-$\sigma$ and $M_{\rm BH}$--$M_\star$ relations (e.g., Bennert \etal\ 2011).
AGN feedback processes are predicted to be most important in the most massive
elliptical galaxies formed within the most massive dark-matter halos.  However,
in the local universe, these same elliptical galaxies contain the oldest
stellar populations and the most dormant SMBHs in terms of their specific mass
accretion rates.  Mass accretion in these systems is likely limited by the
presence of a hot interstellar medium and periodic mechanical feedback from
radio powerful AGN (e.g., Best \etal\ 2005; McNamara \& Nulsen~2012).
Therefore, directly observing the growth of these systems requires challenging
observations of very distant precursors to today's ellipticals.  From models of
large-scale structure formation, it is predicted that growth of modern
ellipticals takes place in high-density regions at \hbox{$z \simgt$~2--3}
(e.g., Kauffmann \etal\ 1996; Governato \etal\ 1998; Volonteri \etal\ 2003;
De~Lucia \etal\ 2006).

Thus far, efforts to observe the distant $z \approx$~2--3 galaxy protoclusters
SSA22 and HS1700+64, likely pre-cursors to rich local clusters like Coma (e.g.,
Steidel \etal\ 1998, 2005), have revealed that the AGN fraction is a factor of
\hbox{$\approx$2--16} times larger (1$\sigma$ statistical range) in the
protocluster environment compared with the low-density field (e.g., Lehmer
\etal\ 2009a,b; Digby-North \etal\ 2010).  This contrasts with local galaxy
clusters, which contain lower AGN fractions in their cores (e.g., Martini
\etal\ 2009; Ehlert \etal\ 2012).  The enhanced SMBH growth rate in the
protocluster environment is likely linked to the presence of larger gas
reservoirs being fed onto more massive galaxies (e.g., Steidel \etal\ 2005) and
SMBHs, as well as an increase in mergers before the
cluster becomes virialized.  These structures are therefore ideal for observing
and characterizing the interdependent processes involved in the growth of the
most massive SMBHs and galaxies, and follow-up studies of protocluster AGN have
revealed interesting first results.  For example, in the SSA22 protocluster at
$z = 3.09$, many of the AGN are found to be coincident with large-scale
($\sim$100~kpc) Lyman-$\alpha$ emitting nebulae (Geach \etal\ 2009) that may be
indicative of emission-line gas being powered by AGN.  

\begin{table*}
\begin{center}
\caption{{\it Chandra} Point-Source Catalog of the 2QZ~Clus Field}
\begin{tabular}{lccccccccccc}
\hline\hline
   & \multicolumn{2}{c}{Position (J2000)} & \multicolumn{3}{c}{Net Counts} & \multicolumn{3}{c}{Exposure (ks)} & \multicolumn{3}{c}{Count-Rate (10$^{-4}$ cnts s$^{-1}$)}  \\
   & \multicolumn{2}{c}{\rule{1.5in}{0.01in}}  & \multicolumn{3}{c}{\rule{1.4in}{0.01in}} & \multicolumn{3}{c}{\rule{0.7in}{0.01in}} & \multicolumn{3}{c}{\rule{1.2in}{0.01in}}  \\
 \multicolumn{1}{c}{Source ID}  & $\alpha_{\rm J2000}$ & $\delta_{\rm J2000}$ & FB & SB & HB & FB & SB & HB & FB & SB & HB \\
 \multicolumn{1}{c}{(1)} & (2) & (3) & (4)--(6) & (7)--(9) & (10)--(12) & (13) & (14) & (15) & (16)--(18) & (19)--(21) & (22)--(24)  \\
\hline\hline
 1\dotfill &                     10 03 00.7 &                    +00 10 54.1 &          16.4$^{+10.0}_{-8.9}$ &                        $<$20.2 &                        $<$26.0 & 53.5 & 47.7 & 77.2 &            2.2$^{+2.0}_{-1.8}$ &                         $<$4.5 &                         $<$3.7 \\
 2\ldots\ldots\ldots\ldots\ldots\ldots &                     10 03 00.8 &                    +00 11 25.1 &         42.2$^{+14.2}_{-11.4}$ &          21.9$^{+10.7}_{-7.7}$ &                        $<$36.3 & 24.6 & 21.9 & 35.8 &           25.0$^{+6.1}_{-4.9}$ &           13.2$^{+5.2}_{-3.7}$ &                        $<$10.8 \\
 3\dotfill &                     10 03 05.9 &                    +00 09 49.2 &         67.3$^{+12.1}_{-11.0}$ &                        $<$15.5 &         66.8$^{+11.7}_{-10.6}$ & 61.7 & 57.1 & 80.5 &           11.3$^{+2.1}_{-1.9}$ &                         $<$2.8 &            8.8$^{+1.6}_{-1.4}$ \\
 4\dotfill &                     10 03 09.2 &                    +00 11 31.1 &           24.7$^{+8.9}_{-7.7}$ &           14.4$^{+6.4}_{-5.1}$ &                        $<$22.4 & 71.6 & 68.7 & 83.3 &            3.4$^{+1.3}_{-1.1}$ &            2.1$^{+0.9}_{-0.8}$ &                         $<$2.8 \\
 5\dotfill &                     10 03 09.3 &                    +00 12 24.3 &         64.2$^{+11.3}_{-10.2}$ &           27.8$^{+7.6}_{-6.4}$ &           38.3$^{+9.4}_{-8.3}$ & 69.5 & 66.1 & 83.0 &            9.2$^{+1.7}_{-1.5}$ &            4.2$^{+1.2}_{-1.0}$ &            4.6$^{+1.2}_{-1.0}$ \\
 6\dotfill &                     10 03 09.6 &                    +00 09 01.1 &         68.9$^{+11.4}_{-10.2}$ &           54.5$^{+9.5}_{-8.3}$ &                        $<$21.8 & 46.0 & 42.5 & 60.3 &           15.1$^{+2.4}_{-2.2}$ &           13.2$^{+2.2}_{-1.9}$ &                         $<$3.6 \\
 7\dotfill &                     10 03 09.8 &                    +00 13 12.9 &           14.0$^{+8.0}_{-6.8}$ &           14.3$^{+6.2}_{-5.0}$ &                        $<$19.7 & 65.9 & 61.8 & 82.6 &            2.0$^{+1.2}_{-1.0}$ &            2.3$^{+1.0}_{-0.8}$ &                         $<$2.4 \\
 8\dotfill &                     10 03 13.1 &                    +00 13 05.1 &          57.7$^{+10.2}_{-9.1}$ &           29.0$^{+7.3}_{-6.1}$ &           30.4$^{+8.1}_{-6.9}$ & 74.5 & 71.9 & 84.6 &            7.9$^{+1.4}_{-1.3}$ &            4.1$^{+1.0}_{-0.9}$ &            3.6$^{+1.0}_{-0.8}$ \\
 9\dotfill &                     10 03 13.2 &                    +00 10 26.8 &           41.9$^{+9.8}_{-8.7}$ &           21.6$^{+6.9}_{-5.6}$ &           20.1$^{+7.9}_{-6.7}$ & 76.3 & 74.1 & 85.1 &            5.5$^{+1.3}_{-1.1}$ &            2.9$^{+0.9}_{-0.8}$ &            2.4$^{+0.9}_{-0.8}$ \\
10\dotfill &                     10 03 13.9 &                    +00 12 17.5 &           13.7$^{+7.0}_{-5.7}$ &           11.3$^{+5.4}_{-4.2}$ &                        $<$15.6 & 78.7 & 77.0 & 85.9 &            1.7$^{+0.9}_{-0.8}$ &            1.5$^{+0.7}_{-0.6}$ &                         $<$1.9 \\
\hline
\end{tabular}
\end{center}
NOTE.---[{\it A portion of Table~1 is shown here to illustrate content.  All 133 rows and 28 columns are provided in the electronic version.}] Col.(1): \chandra\ source ID for 2QZ~Clus field. Col.(2) and (3): Right ascension ($\alpha_{\rm J2000}$) and declination ($\delta_{\rm J2000}$), respectively.  Right ascension is quoted in units of degrees, minutes, seconds.  Declination is quoted in degrees, arcminutes, and arcseconds. Col.(4)--(12): Net counts ($N$) and $1\sigma$ upper/lower bounds, computed following $\S$2.2, for FB, SB, and HB. Col.(13)--(15): Vignetting-corrected effective exposure times (in kiloseconds) for FB, SB, and HB, respectively. Col.(16)--(24): Vignetting corrected count-rates and $1\sigma$ upper/lower bounds, computed following the methods described in $\S$2.2, for FB, SB, and HB. Col.(25)--(27): X-ray flux in units of ergs~cm$^{-2}$~s$^{-1}$ for the FB, SB, and HB, respectively. Col.(28): Notes related to source (``M'' = Manual photometry required due to image edge; ``H'' = Coincident with $z = 2.23$ HAE; ``Q'' = Coincident with optically identified QSO).
\end{table*}


To expand upon our efforts to understand AGN activity in the high-density
environment at high redshift, we have performed \chandra\ observations of an
overdensity of actively star-forming galaxies and optically-selected QSOs at $z
= 2.23$: the 2QZ~Cluster~1004+00 structure (hereafter, 2QZ~Clus).  2QZ~Clus was
discovered by Matsuda \etal\ (2011) by first searching the 2dF QSO Redshift
Survey (2QZ; Croom \etal\ 2001, 2004) and identifying an overdensity of quasars
within the redshift interval \hbox{$z =$~2.216--2.248}, corresponding to the
H$\alpha$ line in the UKIRT WFCAM H$_2$S1 filter, and then performing narrow
and broad band near-infrared imaging (via the H$_2$S1 $\lambda_c =
2.121$~$\mu$m and $K$-band filters) to identify star-forming active galaxies at
$z = 2.23$ via the H$\alpha$ emission line.  The redshift range of the
H$\alpha$ line corresponding to the 50\% transmission wavelengths of the
H$_2$S1 band is $z = 2.216$--2.248.  These H$\alpha$ emitters (HAEs) therefore
have both redshift identifications and H$\alpha$-based measures of their
star-formation rates (SFRs).  Matsuda et al.  (2011) revealed that the 2QZ~Clus
field contains an overdensity of 22 vigorous star-forming HAEs (H$\alpha$-based
SFR~$\simgt$~14~\sfr\ without any correction for extinction) in a $13.7 \times
13.7$~arcmin$^2$ region ($\approx$$22 \times 22$~comoving~Mpc$^2$).  These
observations probe the tip of the iceberg of the actively star-forming $z =
2.23$ 2QZ~Clus galaxy population with additional contributions from luminous
AGN.  Identification of the lower-SFR and H$\alpha$-obscured population is
still needed to fully characterize the large-scale nature of this structure.

In this paper, we present first results from the \chandra\ observations of
2QZ~Clus.  In $\S$~2, we discuss our analysis methods and provide reduced data
products, including images, exposure maps, and point-source catalogs.  In
$\S$~3, we present the properties of the \xray\ detected sources in the
2QZ~Clus structure, and compare these with an equivalent unbiased sample of $z
= 2.23$ HAEs found in the COSMOS survey field from the High-$z$ Emission Line Survey
(HiZELS; Geach \etal\ 2008; Sobral \etal\ 2009).  In $\S$~4, we compare the
relative growth rates of 2QZ~Clus SMBHs and galaxies.  We further discuss the
nature of the 2QZ~Clus structure itself and put the AGN activity into the context
of the broader $z = 2.23$ population of HAEs found in COSMOS.  Finally, in
$\S$~5, we summarize our results.

The Galactic column densities for 2QZ~Clus and COSMOS are $3.0 \times
10^{20}$~cm$^{-2}$ and $2.5 \times 10^{20}$~cm$^{-2}$, respectively.  All of
the \hbox{X-ray} fluxes and luminosities quoted throughout this paper have been
corrected for Galactic absorption.  In the \xray\ band, we make use of three
bandpasses: \hbox{0.5--2~keV} (soft band [SB]), \hbox{2--7~keV} (hard band
[HB]), and \hbox{0.5--7~keV} (full band [FB]).  Values of $H_0$ = 70~\hbox{km
s$^{-1}$ Mpc$^{-1}$}, $\Omega_{\rm M}$ = 0.3, and $\Omega_{\Lambda}$ = 0.7 are
adopted throughout this paper (e.g., Spergel \etal\ 2003).

%
\section{Observations, reduction, and analysis}
%

Throughout this paper we will be comparing the SMBH growth among HAEs in the
2QZ~Clus structure with HAEs in the COSMOS field.  To do this
effectively, we designed our \chandra\ observations of 2QZ~Clus to be of
comparable depth and quality as the \chandra\ data products that are already
available from the \chandra\ COSMOS survey (\hbox{C-COSMOS}; Elvis \etal\ 2009;
Puccetti \etal\ 2009; Civano \etal\ 2012).  Therefore the \chandra\ analysis
that follows has been customized to produce catalogs and data products that can
be directly compared with those available from \hbox{C-COSMOS}.

\subsection{Data Reduction}

We obtained a $\approx$100~ks \chandra\ exposure consisting of a single
\hbox{$16.9\arcmin \times 16.9\arcmin$} ACIS-I pointing (\chandra\
Obs-ID~13976; taken over the 13th and 14th of January~2012; PI: B.D.~Lehmer)
centered on the 2QZ~Clus region surveyed by Matsuda \etal\ (2011; see
Fig.~1).\footnote{Note that during the observation, the ACIS-S3 chip was on;
however, due to its large off-axis angle and non-coincidence with any $z
\approx 2.23$ sources in the 2QZ~Clus field, we chose to exclude data from the
ACIS-S3 CCD.}  The observation was centered on the aim point coordinates
$\alpha_{\rm J2000} =$~10:03:43.3 and $\delta_{\rm J2000} =$~$+$00:13:26.47 and
was oriented at a roll angle that was 56.9~deg from north.  The total duration
of the observation was 99.6~ks.  For our data reductions, we made use of
{\ttfamily CIAO}~v.~4.4 with {\ttfamily CALDB}~v4.5.0.  We began by
reprocessing our events lists, bringing level~1 to level~2 using the script
{\ttfamily chandra\_repro}.  The {\ttfamily chandra\_repro} script runs a
variety of {\ttfamily CIAO} tools that identify and remove events from bad
pixels and columns, and filter the events list to include only good time
intervals without significant flares and non-cosmic ray events corresponding to
the standard \asca\ grade set (\asca\ grades 0, 2, 3, 4, 6).  

Using the reprocessed level~2 events list, we constructed a first FB image and
a point-spread function (PSF) map (using the tool {\ttfamily mkpsfmap}), which
corresponded to a monochromatic energy at 1.497~keV and an encircled counts
fraction (ECF) set to 0.393.  We constructed an initial source catalog by
searching our FB image with {\ttfamily wavdetect} (run with our PSF map), which
was set at a conservative false-positive probability threshold of $1 \times 10^{-7}$ and
run over seven scales from 1--8 (using a $\sqrt{2}$ sequence: 1, $\sqrt{2}$, 2,
2$\sqrt{2}$, 4, 4$\sqrt{2}$, and 8).  To sensitively measure whether any
significant flares remained in our observations, we constructed
point-source-excluded \hbox{0.5--7~keV} background light curves for the
observation in a variety of time bins (spanning \hbox{10--800~s} bins).  We
found no evidence of any significant ($\simgt$5~$\sigma$) flaring events
throughout the observation, and therefore considered our reprocessed level 2
events file to be sufficiently cleaned. 

Next, using the initial \xray\ source catalog and an optical $I < 22$~mag
source catalog from the Sloan Digital Sky Survey Data Release 6 (SDSS-DR6;
Adelman-McCarthy \etal\ 2008), we registered our aspect solution and events
list to the SDSS-DR6 frame using {\ttfamily CIAO} tools {\ttfamily
reproject\_aspect} and {\ttfamily reproject\_events}, respectively.  The
resulting astrometric reprojections gave very small astrometric adjustments,
including linear translations of \hbox{$\delta x = -0.12$}~pixels and
\hbox{$\delta y = +0.30$}~pixels, a rotation of $-0.0097$~deg, and a pixel
scale stretch factor of 1.00018.

\subsection{Point-Source Catalog Production}

Using the reprojected aspect solution and events file (see $\S$2.1), we
constructed standard-band images, 90\% ECF PSF maps (corresponding to 1.497~keV,
4.51~keV, and 2.3~keV for the SB, HB, and FB, respectively), and exposure maps
in the three standard bands.  The exposure maps were constructed following the
basic procedure outlined in $\S$~3.2 of Hornschemeier \etal\  (2001); these
maps were normalized to the effective exposures of sources located at the aim
points. This procedure takes into account the effects of vignetting, gaps
between the CCDs, bad column and pixel filtering, and the spatially dependent
degradation of the ACIS optical blocking filter.  A photon index of $\Gamma =
1.4$ was assumed in creating the exposure maps.  In Figure~1, we show the
adaptively-smoothed FB image and the locations of HAEs and QSOs that are
associated with the 2QZ~Clus structure.  The image was smoothed using
{\ttfamily csmooth} and was ``flattened'' by dividing the smoothed image by a
smoothed exposure map.

We constructed a \chandra\ source catalog by (1) searching the three
standard-band images using {\ttfamily wavdetect} (including the appropriate
90\% ECF PSF and exposure maps) at false-positive probability threshold of $2
\times 10^{-5}$ (the equivalent detection threshold adopted by Elvis \etal\
2009 in the \hbox{C-COSMOS} field); and (2) adjoining the three bandpass catalogs
using cross-band matching radii of 2.5~arcsec and 4.0~arcsec for sources with
off-axis angles of $<$6~arcmin and $>$6~arcmin, respectively.  Through this
procedure, we identified a total of 133 unique \xray\ detected sources in the
2QZ~Clus field.  These 133 sources constitute our main \chandra\ catalog (see
Table~1).

%
%
\begin{figure}
\figurenum{2}
\centerline{
\includegraphics[width=9.0cm]{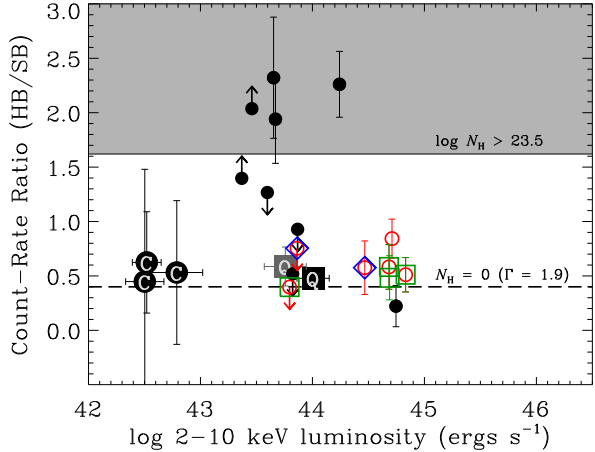}
}
\caption{
Observed X-ray spectral shapes (HB-to-SB count-rate ratio)
vs.~\hbox{2--10~keV} luminosity for \xray\ detected $z = 2.23$ HAEs and QSOs
in 2QZ~Clus and \hbox{C-COSMOS} fields.  The red open circles and green open
squares show 2QZ~Clus HAEs and $z = 2.23$ QSOs, respectively; blue diamonds
indicate sources that satisfy the $BzK$ star-forming galaxy criteria.
\hbox{C-COSMOS} $z = 2.23$ HAEs are shown as filled circles.  Results from
stacking HAE samples in different overdensity bins in C-COSMOS are shown as
black circles with the letter ``C'' and stacking of 2QZ~Clus HAE samples
including and excluding pre-selected QSOs used to identify the 2QZ~Clus structure are
respectively shown as black and gray squares with the letter ``Q'' (see $\S$4.2
for details).  The horizontal dashed line shows the expected count-rate ratio
for an unobscured AGN with a power-law SED of $\Gamma = 1.9$.  The shaded
region shows the expected count-rate ratio for AGN obscured by a column of
$\log (N_{\rm H}/{\rm cm^{-2}}) \simgt 23.5$.  All of the 2QZ~Clus sources have
spectral shapes consistent with luminous unobscured AGN, while the
\hbox{C-COSMOS} AGN span a broader range of spectral slopes.
}
\end{figure}

We performed aperture photometry for our 133 sources in each of the three
standard bands.  For each \xray\ source, we extracted source plus background
counts $S_{\rm src}$ and exposure-map values $T_{\rm src}$ using circular
apertures with radii corresponding to the 90\% ECF, which were determined using
the corresponding PSF-map value at the source location.  We visually inspected
these regions and found that only two source-pairs had very minor overlaps in
the HB PSF (and no overlaps in the smaller SB and FB PSFs).  The overlapping
regions of these source pairs contained $\le$1 HB counts; therefore, we did {\it
not} perform any corrections to the photometry for these sources.

For each source, local background counts and exposure values were then
extracted from each of the three bandpasses. This was achieved by creating
images and exposure maps with all sources masked out of circular masking
regions of radius 1.5 times the size of the 95\% encircled energy fraction
(estimated using each local PSF).  Using these masked data products, we then
extracted background counts $S_{\rm bkg}$ and exposure values $T_{\rm
bkg}$ from a larger background extraction square aperture that was centered
on the source.  The size of the background extraction square varied with each
source and was chosen to contain $\simgt$30--100 background counts in all
bands.  For each bandpass, net source counts $N$ were computed following $N =
(S_{\rm src} - S_{\rm bkg} T_{\rm src}/T_{\rm bkg})/\gamma_{\rm ECF}$, where
$\gamma_{\rm ECF}$ is the ECF appropriate for the source extraction region and
bandpass.  We computed 1$\sigma$ level Poisson errors on the net counts
following the methods described in Gehrels~(1986).

To calculate vignetting-corrected count-rates, we made use of count-rate maps,
which were constructed by dividing the images by the exposure maps.  The
advantage of count-rate maps is that they allow for an accurate accounting of
the source intensity when gradients in the exposure are present in a source
extraction region (e.g., near chip gaps, image edges, and bad pixels).  We made
use of the same source and background regions described above to extract
on-source count-rates $\phi_{\rm src}$ and background count-rates $\phi_{\rm
bkg}$.  Net count-rates were then computed following \hbox{$\phi = (\phi_{\rm
src} - \phi_{\rm bkg} A_{\rm src}/A_{\rm bkg})/\gamma_{\rm ECF}$}, where
$A_{\rm src}$ and $A_{\rm bkg}$ are the source and background extraction areas
that contain exposure.

For each source, we converted our vignetting-corrected count-rates to fluxes
using conversion factors of [1.18, 0.659, and 1.96]~$\times
10^{-11}$~ergs~cm$^{-2}$~s$^{-1}$~(cnts~s$^{-1}$)$^{-1}$ for the FB, SB, and
HB, respectively.  These factors assume a power-law SED with $\Gamma = 1.4$
that is corrected for Galactic extinction.  In Table~1, we provide the basic
\xray\ properties of the 133 main-catalog sources in the 2QZ~Clus field.  The
survey reaches ultimate 5-count sensitivity limits of $\approx$6.0~$\times
10^{-16}$~\flux, $\approx$3.4~$\times 10^{-16}$~\flux, $\approx$1.0~$\times
10^{-15}$~\flux\ for the FB, SB, and HB, respectively; at $z = 2.23$, these
limits allow for the detection of a source with rest-frame \hbox{2--10~keV}
luminosity $L_{\rm X} \simgt 10^{43}$~\lum.  These limits are nicely compatible with
those achieved by the \hbox{C-COSMOS} survey, which reaches a factor of
\hbox{$\approx$1.4--1.7} times deeper in ultimate sensitivity (Elvis \etal\
2009).

%
%
\begin{figure}[t]
\figurenum{3}
\centerline{
\includegraphics[width=9.0cm]{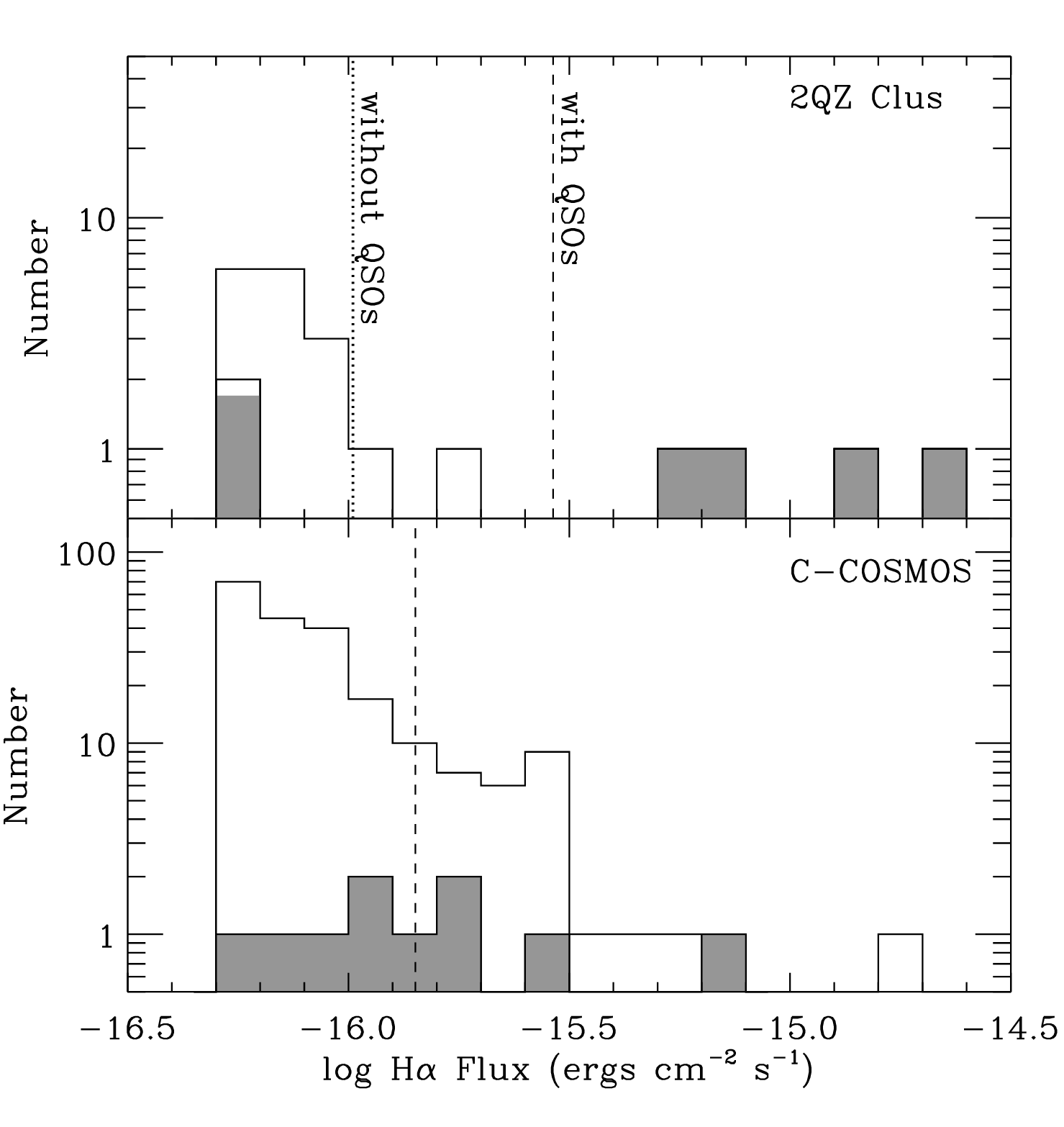}
}
\caption{
H$\alpha$ flux distributions of $z = 2.23$ HAEs ({\it open histograms\/}) and
their AGN subpopulations ({\it filled histograms\/}) for the 2QZ~Clus ({\it
top\/}) and \hbox{C-COSMOS} ({\it bottom\/}) samples.  The AGN population
appears to be preferentially located in HAEs with high H$\alpha$ flux
indicating that the HAE flux itself is likely influenced by the AGN.  The
vertical dashed line in each panel shows the mean H$\alpha$ flux of each
sample.  We note that the mean H$\alpha$ flux of 2QZ~Clus including pre-selected QSOs appears to be higher
than any non-AGN HAEs, suggesting that the selection of this sample is overwhelmed by the 
AGN.  When we exclude pre-selected QSOs, the mean H$\alpha$ flux becomes more
consistent with that of the C-COSMOS field and the star-forming galaxy
population ({\it vertical dotted line\/}).  
}
\end{figure}

\begin{table*}
\begin{center}
\caption{X-ray Detected HAEs and QSOs in the 2QZ~Clus and C-COSMOS Fields}
\begin{tabular}{ccccccccccc}
\hline\hline
\multicolumn{2}{c}{Position (J2000)} & & \multicolumn{2}{c}{Count Rate (10$^{-4}$ cnts~s$^{-1}$)} &  &  &  &  &  & \\
\multicolumn{2}{c}{\rule{1.2in}{0.01in}}  & & \multicolumn{2}{c}{\rule{1.2in}{0.01in}} &  & $\log f_{\rm 0.5-7~keV}$ & $\log L_{\rm X}$ & $\log f_{\rm H\alpha}$ & $\log L_{\rm H\alpha}$  &  \\
$\alpha_{\rm J2000}$ & $\delta_{\rm J2000}$ & XID & SB & HB & $\phi_{\rm 2-7~keV}/\phi_{\rm 0.5-2~keV}$ & (ergs cm$^{-2}$ s$^{-1}$) & (ergs s$^{-1}$) & (ergs cm$^{-2}$ s$^{-1}$) & (ergs s$^{-1}$) & Notes \\
 (1) & (2) & (3) & (4) & (5) & (6) & (7) & (8) & (9) & (10) & (11) \\
\hline\hline
\multicolumn{11}{c}{2QZ~Clus}\\
\hline
    10 03 23.0 &    +00 07 25.0 &       17 &                 15.1~$\pm$~1.8 &                  7.3~$\pm$~1.2 &                0.48 $\pm$ 0.20 &                        $-$13.6 & 44.7 &                         \ldots &                         \ldots &                            QSO \\
    10 03 38.3 &    +00 18 23.8 &       49 &                 13.0~$\pm$~1.6 &                 10.9~$\pm$~1.5 &                0.84 $\pm$ 0.18 &                        $-$13.5 & 44.7 &                        $-$15.3 &                           43.3 &                                \\
    10 03 39.8 &    +00 21 10.8 &       58 &                 14.6~$\pm$~1.8 &                  8.5~$\pm$~1.4 &                0.58 $\pm$ 0.21 &                        $-$13.6 & 44.7 &                        $-$14.8 &                           43.7 &                            QSO \\
    10 03 51.6 &    +00 15 02.1 &       96 &                  3.0~$\pm$~1.1 &                         $<$1.2 &                        $>$0.40 &                        $-$14.5 & 43.8 &                        $-$15.2 &                           43.4 &                            QSO \\
    10 04 04.3 &    +00 14 14.4 &      118 &                  8.7~$\pm$~1.3 &                  5.0~$\pm$~1.0 &                0.58 $\pm$ 0.25 &                        $-$13.8 & 44.5 &                        $-$16.3 &                           42.3 &                          $BzK$ \\
    10 04 09.1 &    +00 15 45.9 &      126 &                  2.7~$\pm$~0.8 &                         $<$2.0 &                        $>$0.75 &                        $-$14.4 & 43.9 &                        $-$16.2 &                           42.3 &                          $BzK$ \\
    10 04 12.9 &    +00 12 57.9 &      129 &                 21.3~$\pm$~1.9 &                 10.9~$\pm$~1.4 &                0.51 $\pm$ 0.16 &                        $-$13.4 & 44.8 &                        $-$14.7 &                           43.9 &                            QSO \\
\hline
\multicolumn{10}{c}{C-COSMOS}\\
\hline
    09 58 54.6 &    +02 14 03.6 &      864 &                  2.3~$\pm$~0.6 &                  5.2~$\pm$~0.9 &                2.26 $\pm$ 0.30 &                        $-$14.0 &                           44.2 &                        $-$16.2 &                           42.4 &                                \\
    10 00 02.6 &    +02 19 58.7 &      774 &                  0.7~$\pm$~0.2 &                  1.3~$\pm$~0.3 &                1.94 $\pm$ 0.41 &                        $-$14.6 &                           43.7 &                        $-$15.6 &                           43.0 &                          $BzK$ \\
    10 00 26.6 &    +01 58 23.0 &      716 &                  0.6~$\pm$~0.3 &                  1.3~$\pm$~0.4 &                2.32 $\pm$ 0.56 &                        $-$14.6 &                           43.7 &                        $-$15.9 &                           42.7 &                          $BzK$ \\
    10 00 44.2 &    +02 02 06.9 &     1456 &                         $<$0.5 &                  1.1~$\pm$~0.3 &                        $<$2.04 &                        $-$14.8 &                           43.5 &                        $-$16.0 &                           42.6 &                          $BzK$ \\
    10 00 55.4 &    +01 59 55.4 &     1486 &                         $<$0.6 &                  0.8~$\pm$~0.3 &                        $<$1.40 &                        $-$14.9 &                           43.4 &                        $-$16.1 &                           42.5 &                          $BzK$ \\
    10 00 55.4 &    +02 33 30.1 &     1386 &                  2.8~$\pm$~0.6 &                         $<$2.6 &                        $>$0.93 &                        $-$14.4 &                           43.9 &                        $-$15.8 &                           42.8 &                                \\
    10 00 57.5 &    +02 33 45.2 &      929 &                  2.7~$\pm$~0.6 &                         $<$1.4 &                        $>$0.52 &                        $-$14.4 &                           43.8 &                        $-$15.7 &                           42.9 &                                \\
    10 01 15.2 &    +01 57 13.3 &     1105 &                  1.5~$\pm$~0.3 &                         $<$1.9 &                        $>$1.27 &                        $-$14.7 &                           43.6 &                        $-$16.2 &                           42.4 &                          $BzK$ \\
    10 01 29.9 &    +02 17 04.1 &     1261 &                  0.5~$\pm$~0.2 &                         $<$1.7 &                        $>$3.21 &                        $-$15.1 &                           43.2 &                        $-$15.1 &                           43.5 &                                \\
    10 01 53.5 &    +02 11 51.6 &      451 &                 18.0~$\pm$~1.4 &                  4.0~$\pm$~0.7 &                0.22 $\pm$ 0.19 &                        $-$13.5 &                           44.7 &                        $-$16.0 &                           42.6 &                                \\
\hline
\end{tabular}
\end{center}
NOTE.---Col.(1) and (2): Optical/near-IR right ascension ($\alpha_{\rm J2000}$) and declination ($\delta_{\rm J2000}$), respectively.  Right ascension is quoted in units of degrees, minutes, seconds.  Declination is quoted in degrees, arcminutes, and arcseconds. Col.(3): X-ray catalog ID (XID).  For sources in the 2QZ~Clus field, the XID is from this work (see Table~1), and for the C-COSMOS sources, the XID is from Elvis \etal\ (2009).  Col.(4) and (5): Count-rate for the SB and HB, respectively, in units of $10^{-4}$~cnts~s$^{-1}$. Col.(6): Count-rate ratio between the 2--7~keV and 0.5--2~keV bands. Col.(7): Logarithm of the \hbox{0.5--7~keV} source flux in ergs~cm$^{-2}$~s$^{-1}$. Col.(8): Logarithm of the rest-frame 2--10~keV luminosity in units of ergs~s$^{-1}$.  Luminosities were calculated  assuming each source is at $z = 2.23$. Col.(9) and (10): Logarithm of the H$\alpha$ source flux (in ergs~cm$^{-2}$~s$^{-1}$) and luminosity (in ergs~s$^{-1}$), respectively, as provided by Matsuda \etal\ (2011) for the 2QZ~Clus field and Sobral \etal\ (2012) for the C-COSMOS field. Col.(11): Notes on \xray\ detected sources.
\end{table*}


%
\section{Results}
%

\subsection{AGN Activity within $z = 2.23$ 2QZ~Clus Structure}

To measure the AGN activity in the 2QZ~Clus structure at $z = 2.23$, we made
use of the catalog of 22 HAEs from Matsuda \etal\ (2011) and four known $z =
2.23$ QSOs.  After accounting for overlap in the HAE/QSO populations, there are
a total of 23 unique $z = 2.23$ source candidates (see Fig.~1).  The one QSO
that does not have an HAE counterpart lies outside the footprint of the
narrow-band survey and is likely to be an HAE itself based on its selection
from optical emission lines.  In the HAE survey, Matsuda \etal\ (2011) selected
candidate $z = 2.23$ sources that satisfied the following criteria: (1)
$K-$H$_2$S1~$\ge 0.251$ (EW$_{\rm obs} \ge 50$~\AA) and (2) the H$\alpha$ flux
excess significance (i.e., the ratio between H$_2$S1 excess and the uncertainty
in the $K-$H$_2$S1 color) $\Sigma \ge 2.5$.  The narrow-band imaging reaches a
H$_2$S1 $5\sigma$ depth of $\approx$19.9~mag (AB).  This selection is
inherently subject to minor (at $\simlt$10--20\% level) contamination from
Pa$\alpha$ emitters at $z = 0.13$, Pa$\beta$ emitters at $z = 0.65$, and
[O~{\small III}]5007 emitters at $z = 3.24$; however, for many cases we can
make use of $B-z^\prime$ and $z^\prime-K$ colors (i.e., the $BzK$ criteria) to
constrain the redshifts to \hbox{$z \sim 1.4$--2.5} and rule-out such
contaminants.  The $BzK$ criteria includes ($z^\prime-K$)$_{\rm AB} -
$($B-z^\prime$)$_{\rm AB} \ge -0.2$ or ($z^\prime-K$)$_{\rm AB} > 2.5$ (Daddi
\etal\ 2004).  From Matsuda \etal\ (2011), 17 out of the 23 $z = 2.23$ are
either spectroscopically confirmed 2QZ~Clus structure member QSOs or satisfy
the $BzK$ criteria (see Fig.~1).  Interestingly, the three QSOs with $B$, $z$,
$K$ photometry do not satisfy the $BzK$ selection criteria, indicating that
AGN-dominated SEDs can be missed by these criteria.  This is likely due to the
presence of strong emission lines, and indeed the three QSOs lie only
$\approx$0.1--0.2~mag away from the $BzK$ selection line (see Fig.~3 of Matsuda
\etal\ 2011).

One HAE of the 23 $z = 2.23$ sources is located outside of the \chandra\ field
of view.  We matched the remaining 22 sources (21 HAEs and four $z = 2.23$
QSOs) to our 2QZ~Clus \chandra\ catalog using a 2\farcs5 matching radius (see
annotations in Fig.~1).  Given the HAE and \xray\ source densities, we estimate
that this choice of matching radius will lead to a negligible number of random
associations (i.e., a total of $\approx$0.02 false-matches expected).  We find
a total of 7 matches, giving an initial AGN fraction of
$\approx$$32^{+17}_{-12}$\% (1$\sigma$ intervals based on Gehrels~1986).  Four
of the seven \xray\ detected sources are the optically-identified QSOs that
were used to originally select the 2QZ~Clus for study (see Matsuda \etal\ 2011
for details); therefore, three additional $z = 2.23$ AGN are identified here as
a result of our \chandra\ observations.  If we exclude the four pre-selected
QSOs, the AGN fraction of the underlying HAE population is
$\approx$$17^{+16}_{-9}$\%.  Six of the 7 \xray\ detected sources are either
spectroscopically confirmed 2QZ~Clus member QSOs or $BzK$ galaxies and are
unlikely to be interlopers.  The one HAE (XID = 49) that does not satisfy
either criteria is inferred to be very luminous $\log L_{\rm X}/{\rm
ergs~s^{-1}} \approx 44.7$ and may have an AGN-dominated SED that does not
satisfy the $BzK$ criteria (similar to the other known QSOs); however, we
cannot rule out the possibility that this source is a low-redshift interloper.
In Figure~2, we show the HB-to-SB count-rate ratio versus \hbox{2--10~keV}
luminosity for \xray-detected HAEs and $z = 2.23$ QSOs in the 2QZ~Clus field,
and in Table~2, we list the properties of these sources.  These seven \xray\
sources have HB-to-SB count-rate ratios and \xray\ luminosities that are
consistent with powerful unobscured QSOs.  

In Figure~3, we show the distributions of H$\alpha$ fluxes for the 2QZ~Clus
HAEs and highlight the subset of HAEs with \xray\ detections.  We find that the
AGN occupy the bright-end of the H$\alpha$ flux distribution ($f_{\rm H\alpha}
\simgt 2 \times 10^{-16}$~\flux), indicating (not surprisingly) that the AGN
themselves have a dominant contribution to the intensity of the H$\alpha$
emission lines in these systems.  The mean H$\alpha$ flux of the 2QZ~Clus
sample overall appears to be higher than any of the non-AGN HAEs, indicating
that the bulk of the H$\alpha$ power from the 2QZ~Clus population is likely to
be from AGN.  However, once the pre-selected QSOs are removed, the mean
H$\alpha$ flux of the 2QZ~Clus sample is within the range of the star-forming
galaxies.  A K-S test reveals that both the 2QZ~Clus and C-COSMOS H$\alpha$
flux distributions are consistent with each other whether or not pre-selected
QSOs are included.

%
\subsection{Comparison with C-COSMOS}
%

One of our key goals is to compare the AGN activity of $z = 2.23$ HAEs in the
2QZ~Clus structure with that of HAE field samples.  As described in $\S$1, the
2QZ~Clus structure was initially selected as an overdensity of QSOs, and
therefore represents a region biased toward both active SMBH mass accretion and
galaxy stellar growth.  To put into broader context the AGN activity in the
2QZ~Clus structure, we make use of an independent sample of \hbox{$z
=$~2.216--2.248} HAEs from HiZELS in the wide-area $\approx$1.6~deg$^2$ COSMOS
survey field (Sobral \etal\ 2012; see also, Geach \etal\ 2012).  The HiZELS
HAEs were selected using the same telescopes, cameras, and filters as used for
the 2QZ~Clus HAE selection; however, unlike 2QZ~Clus, the COSMOS field was not
selected to have any QSO overdensity at $z = 2.23$.  In total there are 353 $z
= 2.23$ HAE candidates with $f_{\rm H\alpha} \simgt 5 \times 10^{-17}$~\flux\
(complete at the $\approx$90\% level) that satisfy the criteria used to select
the 2QZ~Clus HAEs (i.e., $K-$H$_2$S1~$\ge 0.215$; H$_2$S1~$\le 19.9$~mag).  In
this selection, we do not remove non-$BzK$ galaxies for fair comparisons with
2QZ~Clus HAEs and to avoid $z = 2.23$ AGN being removed (see discussion in
$\S$~3.1 regarding $z = 2.23$ 2QZ~Clus QSOs that do not satisfy $BzK$
selection).  As such, our selection is somewhat more liberal than the final
selection of galaxies presented in Sobral \etal\ (2012), which do include a
$BzK$ filtering (slightly modified), together with another color-color
selection ($UBR$), high quality photometric redshifts, and information on
double/triple line emitters (see Sobral \etal\ 2012 for details).  

Of the 353 HiZELS HAEs, 210 overlap with the footprint of the \chandra-COSMOS
survey (\hbox{C-COSMOS}; Elvis \etal\ 2009; Puccetti \etal\ 2009).
\hbox{C-COSMOS} is a contiguous $\approx$0.92~deg$^2$ \chandra\ survey that
reaches \hbox{70--180~ks} depths (vignetting-corrected) across $\approx$84\% of
the surveyed area.  The combination of the HiZELS HAE and \hbox{C-COSMOS}
surveys therefore constitutes a powerful data set by which direct comparisons
can be made with the 2QZ~Clus data.  We made use of the $BzK$ technique to make
a first-order assessment of the contamination fraction for our HAE sample.  Of
these 210 HAEs, we found that 160 had $B$, $z$, and $K$ photometry that would
allow for a $BzK$ redshift assessment.  Of these 160 HAEs, 132 ($\approx$83\%)
have colors satisfying the $BzK$ color selection.  We note that this $BzK$
fraction is larger than that reported by Sobral \etal\ (2012) for the full HAE
sample.  The difference here is that we have limited our analysis to the
relatively bright HAEs in the HiZELS sample, which has a lower contamination
fraction.  Furthermore, the $BzK$ fraction for our HAEs is consistent with that
found for the 2QZ~Clus HAE sample.  We have tested the effects of excluding all
non-$BzK$ HAEs from our samples and find that this choice has an insignificant
impact on the results presented throughout this paper. 

Using a matching radius of $2.5$~arcsec, we obtain a total of 10 \chandra\
sources matched to the 210 HiZELS HAEs that were within the \hbox{C-COSMOS}
footprint, giving an initial AGN fraction of $4.8^{+2.0}_{-1.4}$\%; a factor of
$6.7^{+4.7}_{-3.2}$ times lower than the initial 2QZ~Clus AGN fraction of
$32^{+17}_{-12}\%$.  Given the QSO pre-selection in the 2QZ~Clus field, this
difference in AGN fraction may not be so surprising.  When we exclude the four
pre-selected QSOs from our computation, we find the 2QZ~Clus AGN fraction is
$\approx$$17^{+16}_{-9}$\% (i.e., 3 AGN out of 18 HAEs); still a factor of
$3.5^{+3.8}_{-2.2}$ times higher than the HiZELS AGN fraction.  If we assume
that the \hbox{C-COSMOS} HAE AGN fraction is representative of the HAE
population in general, we estimate that the binomial probability of detecting 3
or more AGN out of 18 HAEs is $\approx$6\%.

In Figure~2, we show the count-rate ratio vs.~2--10~keV luminosity for the
HiZELS $z = 2.23$ HAEs that are \xray\ detected in \hbox{C-COSMOS}.  The HiZELS
HAEs host an almost equal blend of moderately obscured ($N_{\rm H} \simgt
10^{22}$~cm$^2$) and unobscured AGN as measured by their band ratios.  This
appears to differ somewhat from the ``softer'' AGN found in 2QZ~Clus (see
Fig.~2).  If we assume that the $\approx$50\% ratio of obscured to unobscured
AGN in the C-COSMOS HAEs is typical of $z = 2.23$ HAEs, then the probability of
finding all three uniquely \xray\ selected AGN (i.e., excluding the four
pre-selected QSOs) in 2QZ~Clus to be unobscured is $\approx$25\%.  We therefore
do not conclude that the differences between the \xray\ spectral slopes in 2QZ~Clus
HAEs are significantly different from those of \hbox{C-COSMOS} HAEs.  In Figure~3, we
display the H$\alpha$ flux distribution of the \hbox{C-COSMOS} HAEs and
highlight the subset that are \xray\ AGN.  In contrast to 2QZ~Clus HAEs, we
find that the mean C-COSMOS HAE flux is within the distribution of non-AGN
HAEs, indicating that the total H$\alpha$ power output is likely dominated by
the star-forming galaxy population.

\begin{table*}
\begin{center}
\caption{X-ray Stacking of HAE Samples}
\begin{tabular}{cccccccccc}
\hline\hline
 & &  &  &  &  &  $\log L_{\rm X}$ & $\dot{M}_{\rm BH}$ & $S_{250\mu m}$ & SFR  \\
Sample & $\log (\rho/\langle \rho \rangle)$ & $N_{\rm gal}$ & $N_{\rm AGN}$ & $f_{\rm AGN}$  & $\phi_{\rm 2-7~keV}/\phi_{\rm 0.5-2~keV}$ &  (ergs~s$^{-1}$) & ($10^{-3} M_\odot$~yr$^{-1}$) & (mJy) & ($M_\odot$~yr$^{-1}$)  \\
 (1) & (2) & (3) & (4) & (5) & (6) & (7) & (8) & (9) & (10) \\
\hline\hline
     C-COSMOS Low Density \ldots\ldots\ldots\ldots &   $-$0.38 $\pm$ 0.38 &    61 &     2 &         0.03$^{+0.04}_{-0.02}$ &                0.62 $\pm$ 0.47 &               42.52 $\pm$ 0.13 &                     11 $\pm$ 4 &                  3.9 $\pm$ 1.6 &                    68 $\pm$ 29 \\
        \phantom{C-COSMOS} Medium Density \dotfill &      0.38 $\pm$ 0.38 &   122 &     5 &         0.04$^{+0.03}_{-0.02}$ &                0.53 $\pm$ 0.66 &               42.79 $\pm$ 0.23 &                    21 $\pm$ 15 &                  3.4 $\pm$ 0.6 &                    60 $\pm$ 10 \\
          \phantom{C-COSMOS} High Density \dotfill &      1.12 $\pm$ 0.38 &    27 &     3 &         0.11$^{+0.11}_{-0.06}$ &                0.45 $\pm$ 1.03 &               42.50 $\pm$ 0.17 &                     11 $\pm$ 5 &                         $<$2.3 &                          $<$42 \\
                  2QZ~Clus including QSOs \dotfill &      0.31 $\pm$ 0.10 &    21 &     6 &         0.29$^{+0.17}_{-0.11}$ &                0.48 $\pm$ 0.10 &               44.01 $\pm$ 0.14 &                  362 $\pm$ 142 &                  6.0 $\pm$ 2.6 &                   111 $\pm$ 73 \\
        \phantom{2QZ~Clus} excluding QSOs \dotfill &      0.31 $\pm$ 0.10 &    18 &     3 &         0.17$^{+0.16}_{-0.09}$ &                0.59 $\pm$ 0.18 &               43.76 $\pm$ 0.19 &                  204 $\pm$ 110 &                  3.6 $\pm$ 1.3 &                    63 $\pm$ 23 \\
\hline
\end{tabular}
\end{center}
NOTE.---Col.(1): Description of $z = 2.23$ HAE sample being stacked. Col.(2): HAE source overdensity, $\rho/\langle \rho \rangle$, computed as the ratio of the local HAE source density $\rho$ to mean HAE density across the entire HiZELS COSMOS survey area $\langle \rho \rangle$ (Sobral \etal\ 2012).  For the shallower 2QZ~Clus HAE sample, we computed $\rho/\langle \rho \rangle$ using bright HAEs in both the 2QZ~Clus and HiZELS COSMOS fields (see $\S$4 for details). Col.(3): Number of galaxies in the relevant stacking sample.  This number reflects exclusion of HAEs that were outside the C-COSMOS footprint, as well as sources that were in the near vicinity (within $\approx$15~arcsec) of unrelated \xray\ detected sources. Col.(4): Number of sources that were detected in the \xray\ band.  Due to the relatively high survey luminosity limits for $z = 2.23$, these sources are expected to be AGN. Col.(5): AGN fraction of sample. Col.(6): Ratio of stacked \hbox{2--7~keV} to \hbox{0.5--2~keV} count-rates $\phi$. Col.(7): Logarithm of the mean \hbox{2--10~keV} luminosity calculated using the stacked \hbox{0.5--2~keV} emission. Col.(8): Estimate of the mean BH accretion rate based on the \xray\ luminosity (see $\S$4 for details and assumptions). Col.(9): Mean 250$\mu$m flux in mJy based on stacking analyses. Col.(10): Mean SFR based on 250$\mu$m flux (see $\S$4.2).
\end{table*}


%
%
\begin{figure*}[t]
\figurenum{4}
\centerline{
\includegraphics[width=9.2cm]{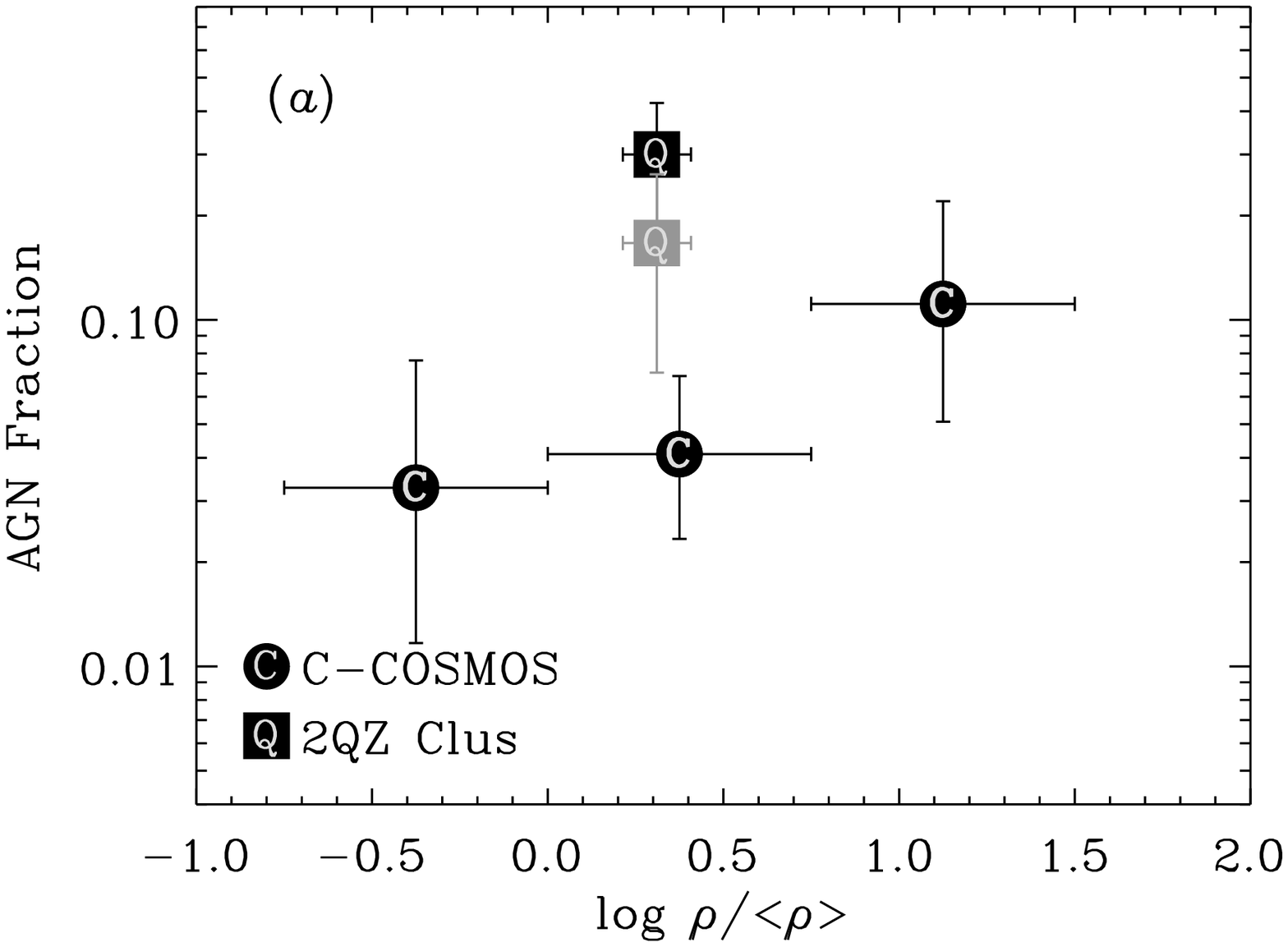}
\hfill
\includegraphics[width=9.2cm]{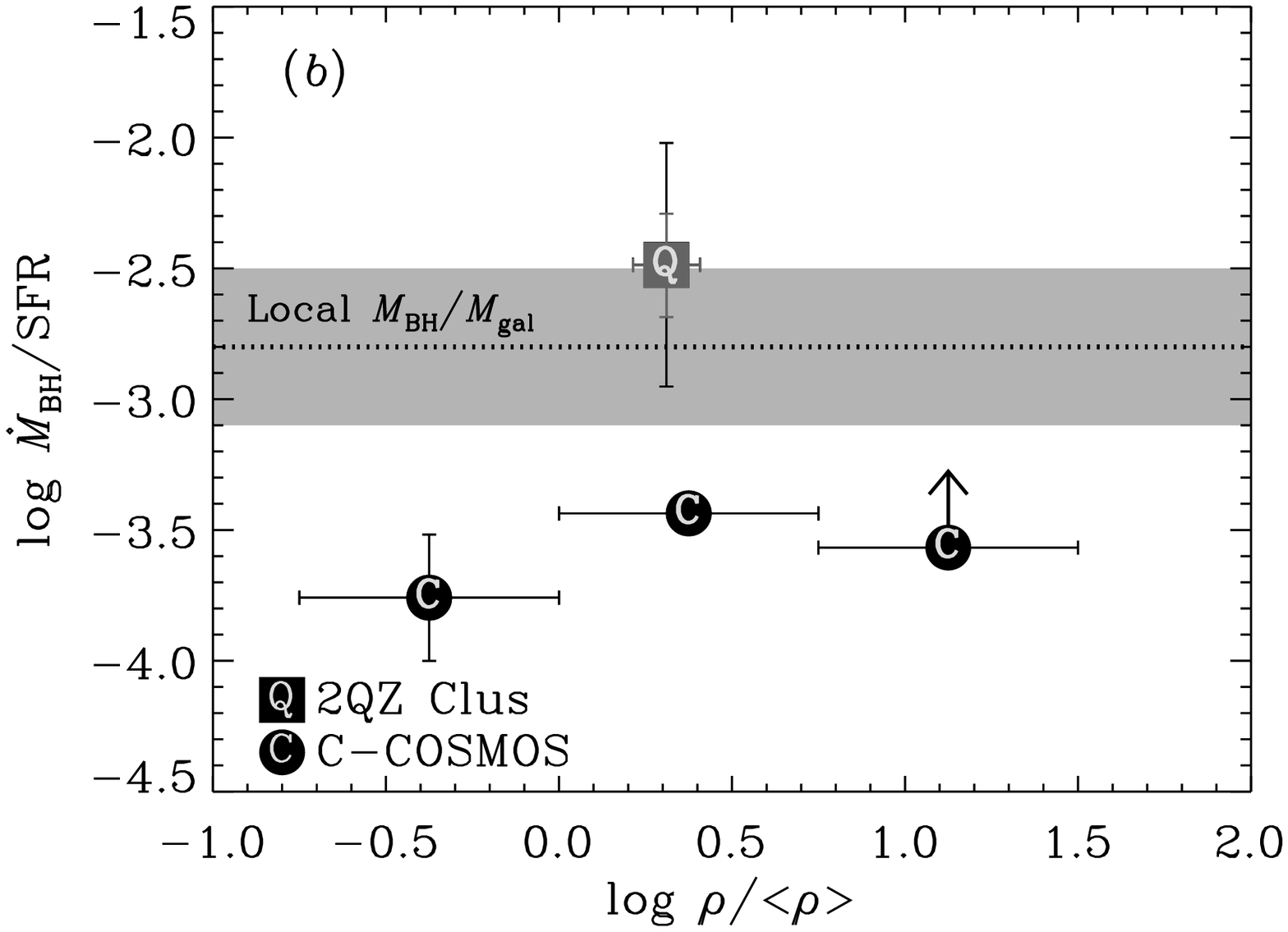}
}
\caption{
($a$) AGN fraction versus local source overdensity for the \hbox{C-COSMOS}
({\it filled circles\/}) and 2QZ~Clus ({\it filled squares\/}) $z = 2.23$ HAEs.
For the 2QZ~Clus field, we have highlighted results for both the inclusion
({\it black squares\/}) and exclusion ({\it gray squares\/}) of the QSOs
that were identified via the 2dF QSO Redshift Survey and used to select the
2QZ~Clus field.  We find suggestive evidence for an increase in AGN fraction
with local density for the \hbox{C-COSMOS} HAEs and find that the 2QZ~Clus AGN
fraction is significantly elevated over that of the \hbox{C-COSMOS} HAEs. ($b$)
Mean SMBH mass accretion rate per SFR vs.~local overdensity for each of the
stacked samples presented in $\S$~4.2 and Table~3.  For comparison, the local
$M_{\rm BH}/M_{\rm gal}$ relation and its dispersion from H{\"a}ring \&
Rix~(2004) has been shown as a dotted line with shaded region.  The
$\dot{M}_{\rm BH}$/SFR values for 2QZ~Clus HAEs appear to be consistent with
the local $M_{\rm BH}/M_{\rm gal}$ relation, while the C-COSMOS HAEs fall below
this relation.  We are unable to determine whether $\dot{M}_{\rm BH}$/SFR
varies with local environment for the C-COSMOS HAEs.
}
\end{figure*}

\section{Discussion}

The above analyses indicate that the 2QZ~Clus HAEs contain an enhanced AGN
fraction compared to HAEs found in the C-COSMOS field.  Here we discuss
the possible circumstances surrounding the enhanced SMBH growth in the 2QZ~Clus
structure.  We further put into context the mutual SMBH and galaxy growth of
2QZ~Clus and C-COSMOS HAEs.

\subsection{Environment as a Driver of Galaxy and SMBH Growth}

To assess the role that environment plays in driving SMBH accretion in the HAE
population, we made use of the wide-area HiZELS HAEs in the \hbox{C-COSMOS}
survey to estimate AGN fraction as a function of source density.  For the
broader HiZELS HAE sample (i.e., the $\approx$1.6~deg$^2$ sample), we measured
local HAE overdensities for each source following $\rho/\langle \rho \rangle
\approx 4/(\pi r_4^2)/\langle \rho \rangle$, where $r_4$ is the separation
between a given source and its fourth nearest neighbor and $\langle \rho
\rangle$ is the mean HAE source density over the entire HiZELS COSMOS field.
Since the HiZELS COSMOS field has wider aerial coverage than the subpopulation
of HAEs that lie within the \hbox{C-COSMOS} footprint, we expect that
$\rho/\langle \rho \rangle$ should provide a good estimate of the local source
environment for all \hbox{C-COSMOS} sources without suffering from
uncertainties due to survey edge effects.  Focusing on only the HAEs that were
within the \hbox{C-COSMOS} footprint, we divided the sample into three bins of
source overdensity, $\log \rho/\langle \rho \rangle =$ [$-0.35$, 0.35,
1.12]~$\pm 0.38$.  For each bin, we computed the fraction of HAEs hosting
luminous AGN detected in the X-ray band.  Figure~4$a$ shows the AGN fraction as
a function of overdensity $\rho/\langle \rho \rangle$ for the \hbox{C-COSMOS}
sources, and in Table~4 we tabulate the AGN fractions of each subset.  We find
suggestive evidence for an increase in AGN fraction with local HAE source
density, such that the $f_{\rm AGN}(\log \rho/\langle \rho \rangle \approx
1.12) = 3.4^{+7.0}_{-2.7} \times f_{\rm AGN}(\log \rho/\langle \rho \rangle
\approx -0.4)$.  A similar trend was noted for Lyman-$\alpha$ emitters in the
$z = 3.1$ SSA22 protocluster (Lehmer \etal\ 2009a).

Given that the 2QZ~Clus HAE survey is somewhat shallower and less complete than
the HiZELS COSMOS HAE survey, we cannot directly compute the equivalent HAE
source density all the way down to the H$\alpha$ flux limit.  Therefore, to
estimate an average $\rho/\langle \rho \rangle$ for all sources in 2QZ~Clus, we
computed the survey-wide 2QZ~Clus HAE source density of a highly complete
($\simgt$90\%) subsample using only relatively bright HAEs ($f_{\rm H\alpha} >
6.3 \times 10^{-17}$~\flux), and then compared this with the source density of
relatively bright COSMOS HiZELS HAEs.  The relative bright-source densities
indicate that the 2QZ~Clus HAEs have on average $\langle \rho \rangle_{\rm
2QZ~Clus}^{\rm bright}/\langle \rho \rangle_{\rm HiZELS}^{\rm bright} = 2.0 \pm
0.5$.  This value is somewhat higher than, although consistent with, the
$\approx$$1.5 \pm 0.4$ value found by Matsuda \etal\ (2011) when comparing the
2QZ~Clus HAEs with a smaller HAE control sample.  We note that the relatively
small area in 2QZ~Clus limits us from effectively measuring the local densities
of each galaxy in the way we did for the HiZELS HAEs.  A wider area HAE survey
of the 2QZ~Clus structure would mitigate this limitation.  In Figure~4$a$, we
highlight the AGN fraction of the 2QZ~Clus HAEs, both including and excluding
HAEs that were pre-selected to be QSOs.  After excluding the pre-selected QSOs,
we find that the 2QZ~Clus AGN fraction is a factor of
$\approx$$4.1^{+5.0}_{-2.8}$ times higher than the AGN fraction of
\hbox{C-COSMOS} HAEs in similarly overdense environments.

%
%
\begin{figure*}[t]
\figurenum{5}
\centerline{
\includegraphics[width=9.2cm]{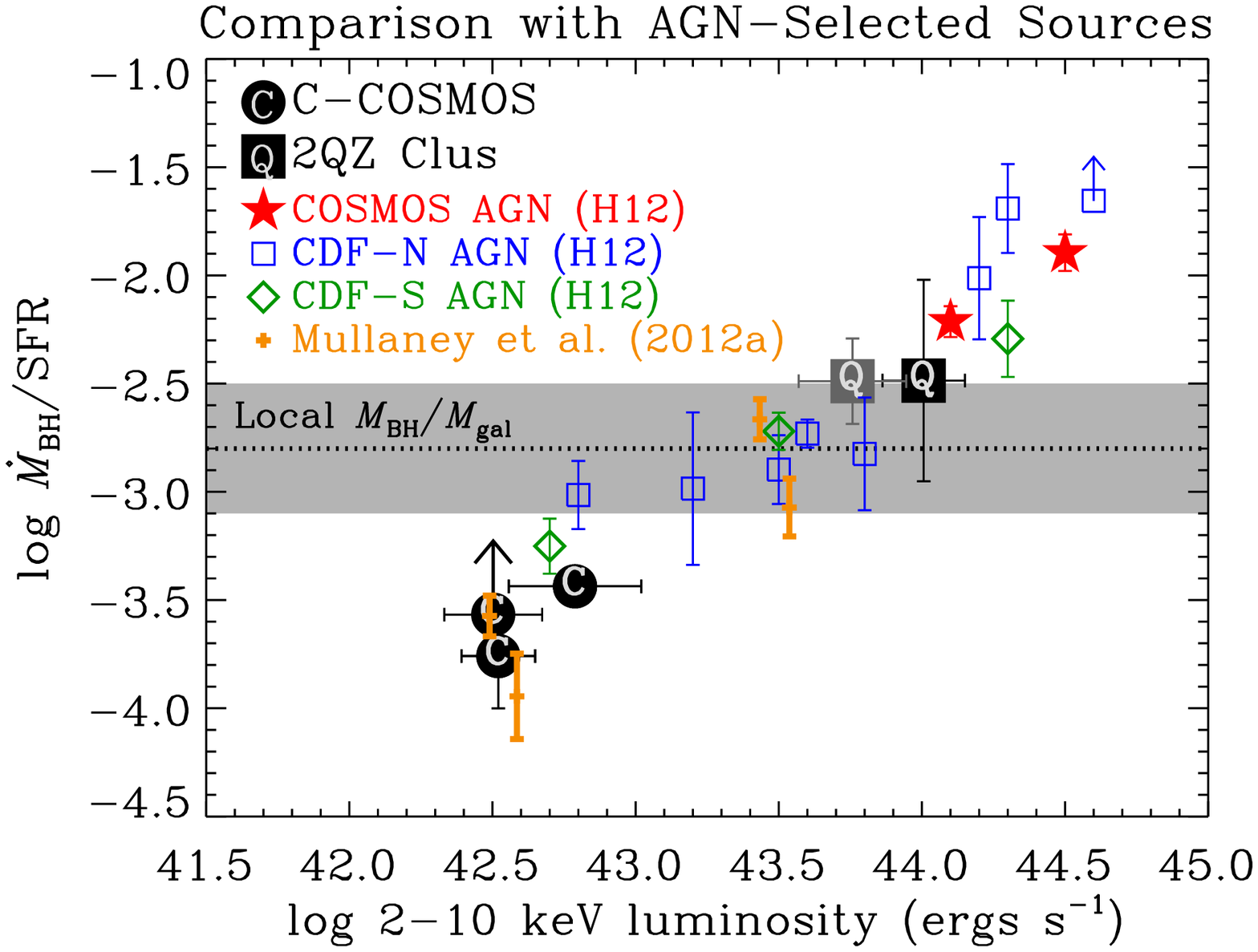}
\hfill
\includegraphics[width=9.2cm]{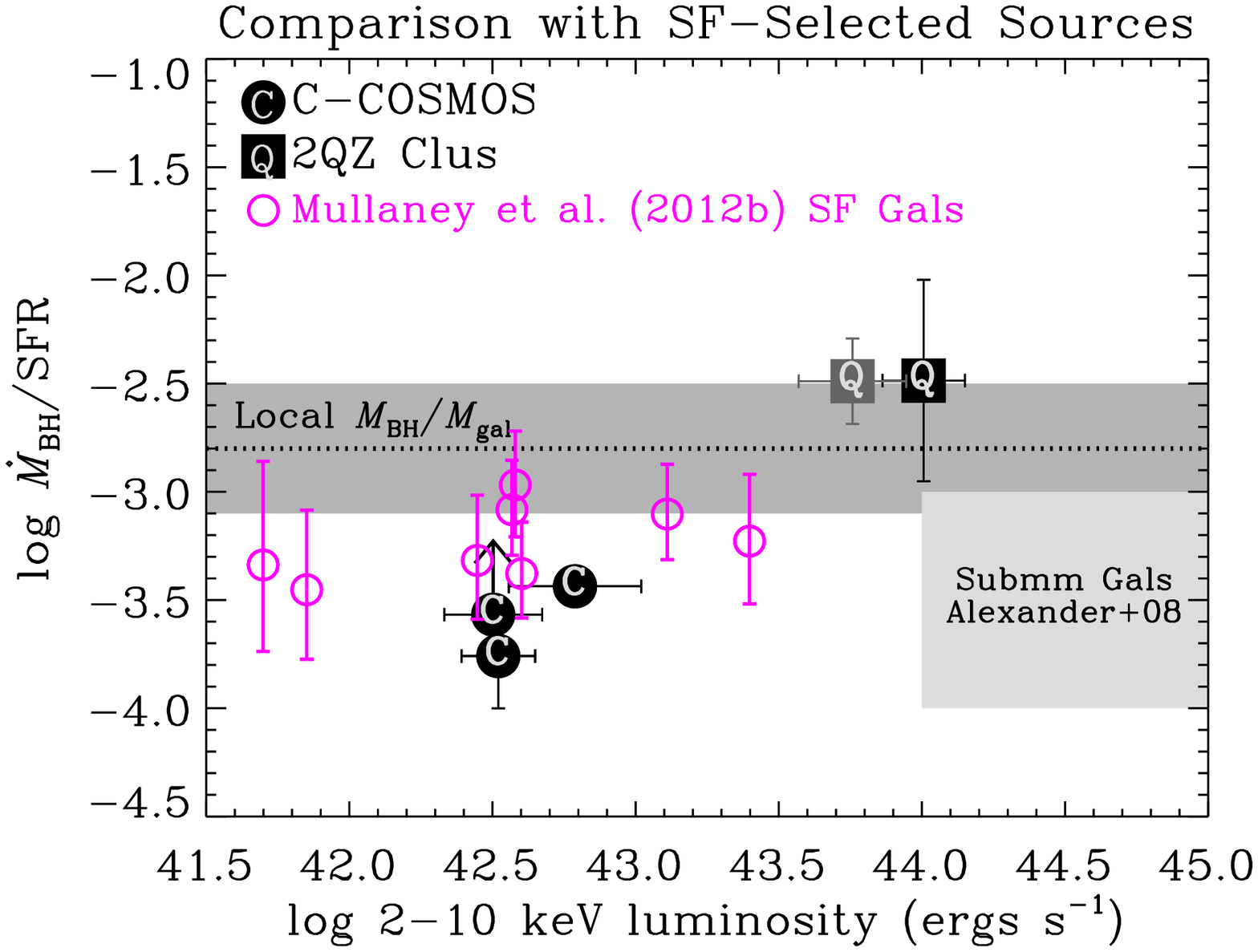}
}
\caption{
Mean SMBH mass accretion rate per SFR ($\dot{M}_{\rm BH}$/SFR) versus
\hbox{2--10~keV} luminosity for the stacked values of 2QZ~Clus and C-COSMOS
HAEs (see Table~3).  In the {\it left panel}, we compare these values with $z
\approx 2$ AGN-selected samples from Harrison \etal\ (2012; H12) and $z
\approx$~1--3 AGN from Mullaney \etal\ (2012a).  In the {\it right panel}, we
compare 2QZ~Clus and C-COSMOS $\dot{M}_{\rm BH}$/SFR values with those of $z
\approx$~1--2 star-forming galaxies from Mullaney \etal\ (2012b) and the range
of vigorously star-forming submillimeter galaxies from Alexander \etal\ (2008).
For reference, we have shown in both panels the local $M_{\rm BH}/M_{\rm gal}$
relation and its dispersion as reported by H{\"a}ring \& Rix~(2004).  The
2QZ~Clus HAEs have mean $\dot{M}_{\rm BH}$/SFR values consistent with the
trends observed for AGN, while the C-COSMOS HAEs appear to be broadly
consistent with both the $z \approx$~1--2 AGN and star-forming galaxy
populations.
}
\end{figure*}

From the above analyses, it appears as though the enhanced AGN fraction in the
2QZ~Clus cannot be explained by an enhanced source density.  However, given
that the HAE source selection is limited to the detection of sources with
high-SFR (SFR~$\simgt 14$~\sfr) or AGN activity, it may be possible that there
are large numbers of 2QZ~Clus galaxies that are lower SFR and/or high H$\alpha$
extinction leading to an inaccurate characterization of the local source
environment.  For this to explain the relatively large AGN fraction in terms of
environment would require that the underlying galaxy population in 2QZ~Clus be
very different from that of C-COSMOS.  Future observations sensitive to these
galaxy populations would be needed to address this.

If we assume that the enhanced AGN fraction in the 2QZ~Clus structure is due
solely to the biased QSO-based selection and that our estimates of the local
source overdensities are representative of the true environments in which the
galaxies are found, we can assess how rarely we should find such elevated AGN
fractions.  From our analysis of the \hbox{C-COSMOS} HAEs, we know that the
probability of an HAE hosting an AGN in an environment like that of 2QZ~Clus is
$\approx$4\%.  Assuming this incidence fraction is representative of the
broader HAE population, we used the binomial probability distribution to
compute the rarity of finding AGN fraction enhancements like those of 2QZ~Clus.
We find that the probability of detecting 3 or more AGN in 18 HAEs (i.e., after
removing pre-selected QSOs) is $\approx$4.1\%.  This suggests that there is
likely to be something inherently ``special'' about the 2QZ~Clus region;
perhaps some unaccounted for physical mechanism is responsible for the enhanced
AGN activity that is unique to 2QZ~Clus.  It is also possible that the elevated
AGN fraction is a product of the presence of more massive host galaxies and
SMBHs.  For example, Xue \etal\ (2010) found that the AGN fraction increases
rapidly with stellar mass.  From their AGN fraction versus stellar mass
relations for luminous AGN (see Fig.~14B of Xue \etal\ 2010), we infer that the
2QZ~Clus galaxies would have to be on-average a factor of $\approx$5--8~times
more massive than the C-COSMOS galaxy population to explain the enhanced AGN
fraction that is observed.  The 2QZ~Clus and C-COSMOS HAEs were selected at the
same SFR limits and have similar mean SFRs (see below).  Since SFR is in
general correlated with stellar mass, with a dispersion of a factor of
$\approx$2 (e.g., Elbaz \etal\ 2007; Salim \etal\ 2007), it is unlikely that
the mean stellar masses of the HAEs in 2QZ~Clus would be $\approx$5--8 times
higher on-average than those in C-COSMOS.  Better characterization of the
2QZ~Clus HAE source properties, large-scale environment, and underlying non-HAE
galaxy population will help us understand this.  This could be achieved via
\hst\ WFC3 infrared observations of the HAE population and/or spectroscopic
follow-up of BX/BM and $BzK$ galaxy candidates.

\subsection{Relative SMBH and Galaxy Growth Rates}

To estimate the mean SMBH growth rates of 2QZ~Clus and C-COSMOS HAEs, we
performed \xray\ stacking following the techniques described in Lehmer \etal\
(2008), with errors measured as 1$\sigma$ intervals of bootstrap resampling
(see, e.g., Basu-Zych \etal\ 2012 for details).  In 2QZ~Clus, we stacked the
total HAE samples both including and excluding the pre-selected QSOs.  For
C-COSMOS, we stacked each of the three subsamples that were divided based on
source density (see $\S$4.1).  In Table~3, we tabulate our stacking results for
the five subsamples.  We find that the stacked \hbox{2--10~keV} luminosities
range from $\log L_{\rm X}/({\rm ergs~s^{-1}}) \approx$~42.5--44, indicating
that AGN clearly dominate the mean stacked \xray\ emission over normal galaxy
emission by more than an order of magnitude for galaxies of similar SFRs and
redshifts (e.g., Lehmer \etal\ 2010; Basu-Zych \etal\ 2012).  In all cases the
\hbox{2--7~keV}/\hbox{0.5--2~keV} count-rate ratios indicate the mean stacked
spectrum is consistent with unobscured AGN (see Fig.~2 and Col.~6 in Table~3).
We converted our mean \hbox{2--10~keV} luminosities into bolometric
luminosities $L_{\rm bol}^{\rm AGN}$ using a correction factor of 22.4, which
corresponds to the median bolometric correction for local AGN with $L_{\rm X}
\approx 10^{41}$--$10^{46}$~\lum\ (Vasudevan \& Fabian~2007).  We then
estimated the mean SMBH accretion rates following $\dot{M}_{\rm BH} \approx
(1-\epsilon)L_{\rm  bol}^{\rm AGN}/(\epsilon c^2)$, where $c$ is the speed of
light and $\epsilon$ is the efficiency by which accreting mass is converted
into radiative energy; here we assume $\epsilon \approx 0.1$ (see, e.g.,
Marconi \etal\ 2004 for motivation).  These assumptions yield mean mass
accretion rates spanning $\dot{M}_{\rm BH} \approx$~0.01--0.02~\sfr\ for the
\hbox{C-COSMOS} HAE samples and $\approx$0.2--0.4~\sfr\ for the 2QZ~Clus HAE
samples (see Col.~8 in Table~3).

The above \xray\ stacking shows that the mean SMBH accretion rates of 2QZ~Clus
HAEs are significantly larger than those of \hbox{C-COSMOS} HAEs; however, we
want to test whether the corresponding SFRs of the HAEs are on average
consistent with that expected from the local $M_{\rm BH}/M_{\rm gal}$ relation.
Although the H$\alpha$ emission line provides an immediate proxy for the
star-formation rate (SFR), H$\alpha$ is often subject to extinction in starburst galaxies
like those studied here (e.g., Calzetti \etal\ 1997).  Furthermore, in the case
of QSOs, the H$\alpha$ emission line fluxes from the AGN will overwhelm any
signal related to star-formation activity, making it difficult to infer
directly the SFRs of these sources (see, e.g., Fig.~3).  

To remedy these issues, we have made use of \herschel\ observations at
250$\mu$m using the Spectral and Photometric Imaging Receiver (SPIRE).
\herschel\ SPIRE data are available in the \hbox{C-COSMOS} field via the HerMES
campaign (Oliver \etal\ 2012) and in the 2QZ~Clus field through a GO program
(PI: Matsuda; Matsuda \etal\ 2013, in-preparation).  These data allow us to
probe emission from the rest-frame 80~$\mu$m, which is well beyond the peak of
the expected dust-emission associated with QSOs ($\approx$6--15~$\mu$m; e.g.,
Efstathiou \& Rowan-Robinson~1995; Netzer \etal\ 2007; Hatziminaoglou \etal\
2010; Mullaney \etal\ 2011).  We therefore anticipate that SPIRE probes the
cool dust emission associated with star-formation activity, and the total SFR
can be measured using these data.  We performed SPIRE stacking of the three
\hbox{C-COSMOS} and two \hbox{2QZ~Clus} samples described in Table~3 following
the techniques detailed in Harrison \etal\ (2012).  We obtained significant
signal in the 250$\mu$m stacks for all samples except for the highest
overdensity C-COSMOS subsample (see Table~3).  Errors on mean SPIRE flux
represent the largest 1$\sigma$ interval based on bootstrap resampling (see
Harrison \etal\ 2012 for details).  We converted 250$\mu$m fluxes to $L_{\rm
IR}$ (8--1000~$\mu$m) using the SED library of Chary \& Elbaz (2001), selecting
a SED (redshifted to $z = 2.23$) on the basis of the total monochromatic
luminosity probed by the 250$\mu$m emission.  We converted the infrared
luminosities to SFRs following \hbox{SFR/(\sfr)~$\approx 9.8 \times 10^{-11}
\times L_{\rm IR}/L_{\odot}$}.  This conversion assumes negligible
contributions from UV emission and is appropriate for a Kroupa~(2001) IMF (see
eqn.~1 of Bell \etal\ 2005).  The mean SFRs are reported in Column~10 of
Table~3.

From the above \xray\ and infrared stacking analyses, we obtain mean
$\dot{M}_{\rm BH}$/SFR ratios for our samples.  In Fig~4$b$, we display
$\dot{M}_{\rm BH}$/SFR versus overdensity ($\rho/\langle \rho \rangle$) for
these samples.  We find that the 2QZ~Clus HAEs have $\dot{M}_{\rm
BH}$/SFR~$\approx$~3.2~$\times 10^{-3}$, similar to the mean $M_{\rm
BH}/M_{\rm gal}$ ratio at $z = 0$ (e.g., H{\"a}ring \& Rix~2004; Bennert \etal\
2011).  For the C-COSMOS HAE subsamples, we find much lower values of
$\dot{M}_{\rm BH}$/SFR~$\approx$~\hbox{(0.2--0.4)}~$\times 10^{-3}$.  Such a
deficit in SMBH-to-galaxy growth rate ratio may be due to the SFR-biased
selection of \hbox{C-COSMOS} HAEs, which has constrained our analysis to
include only powerful star-forming galaxies with extinction uncorrected
H$\alpha$-based SFR$_{\rm H\alpha} > 14$~\sfr.  

Recent \herschel\ SPIRE analyses of \xray\ selected AGN in the CDF-N, CDF-S,
and COSMOS fields have shown that on average the mean SFR increases with mean
AGN luminosity; however, there is some debate as to whether the most powerful
QSOs with $L_{\rm X} \simgt 10^{44}$~\lum\ have declining SFRs (e.g., Harrison
\etal\ 2012; Page \etal\ 2012).  In the left panel of Figure~5, we compare the
2QZ~Clus and C-COSMOS HAE $\dot{M}_{\rm BH}$/SFR values with those of \xray\
selected AGN presented in Harrison \etal\ (2012) and $z \approx$~1--3 AGN from
Mullaney \etal\ (2012a).  \xray\ luminosities have been converted to
$\dot{M}_{\rm BH}$ following the techniques discussed above and SFRs have been
converted to be consistent with our choice of Kroupa~(2001) IMF.  It is
apparent that the mean $\dot{M}_{\rm BH}$/SFR for \xray\ selected AGN appears
to be consistent with the local $M_{\rm BH}/M_{\rm gal}$ relation (H{\"a}ring
\& Rix~2004) for AGN with $L_{\rm X} \simlt 10^{44}$~\lum.  At higher $L_{\rm
X}$, $M_{\rm BH}/M_{\rm gal}$ increases dramatically.  The 2QZ~Clus HAE samples
have $\dot{M}_{\rm BH}$/SFR vs.~$L_{\rm X}$ mean values in good agreement with
the trends observed for the Harrison \etal\ (2012) \xray\ selected AGN (see
Fig.~5).   As discussed above, the C-COSMOS HAEs appear to have somewhat lower
$\dot{M}_{\rm BH}$/SFR values below that expected from the local $M_{\rm
BH}/M_{\rm gal}$ relation; however, these values appear to be consistent with
those of the low-luminosity \xray\ selected AGN from Mullaney \etal\ (2012a).
Given that the mean SFRs and masses of all the AGN samples shown in the
left-panel of Figure~5 are relatively high (SFR~$\approx$~10--100~\sfr\ and
$M_\star \simgt 5 \times 10^{10}$), this trend suggests that SMBHs and galaxies
do not grow simultaneously.  

In the right-panel of Figure~5, we compare the 2QZ~Clus and C-COSMOS HAE
$\dot{M}_{\rm BH}$/SFR values with those of \hbox{$z \approx$~1--2}
star-forming galaxies (Mullaney \etal\ 2012b).  We find that the
\hbox{C-COSMOS} HAEs have $\dot{M}_{\rm BH}$/SFR in good agreement with field
$z \approx$~1--2 star-forming galaxies, which appear to also agree with $z
\approx 2$ submillimeter galaxies studied by Alexander \etal\ (2008).  These
values appear to lie a factor of $\approx$3 below the local $M_{\rm BH}/M_{\rm
gal}$.  We note, that the various assumptions (e.g., bolometric correction and
accretion efficiency) and unknown uncertainties used in our calculations may
contribute to systematic offsets in $\dot{M}_{\rm BH}$/SFR.  However, if our
findings are correct, typical star-forming galaxies must undergo vigorous SMBH
growth in short duty cycles, where $\dot{M}_{\rm BH}$/SFR is above the local
$M_{\rm BH}/M_{\rm gal}$ ratio.  If the QSO phase at $L_{\rm X} \simgt
10^{44}$~\lum\ is responsible for such episodic growth, we can use our
observations to estimate the QSO duty cycle required to bring star-forming
galaxies onto the $M_{\rm BH}/M_{\rm gal}$ relation:
\begin{equation}
M_{\rm BH}/M_{\rm gal} \approx (\dot{M}_{\rm BH}/{\rm SFR})_{\rm QSO}f_{\rm
QSO} + (\dot{M}_{\rm BH}/{\rm SFR})_{\rm gal}(1 - f_{\rm QSO}),
\end{equation}
where $f_{\rm QSO}$ is the fraction of time that a galaxy is in a QSO phase.
Taking $M_{\rm BH}/M_{\rm gal} =$~$10^{-2.5}$--$10^{-3.0}$, ($\dot{M}_{\rm
BH}$/SFR)$_{\rm QSO} \approx 10^{-1.8}$, and ($\dot{M}_{\rm BH}$/SFR)$_{\rm
gal} \approx 10^{-3.3}$, we find \hbox{$f_{\rm QSO} \approx$~2--8\%}.  For
\hbox{C-COSMOS} HAEs, we estimate that the QSO fraction for $L_{\rm X} \simgt
10^{44}$~\lum\ AGN is $\approx$1.0$^{+1.3}_{-0.6}$\%.  This value and our
duty-cycle estimate of $f_{\rm QSO}$ are consistent within errors, indicating
that the short-term SMBH growth in a QSO phase may be sufficient to bring the
typical HAE up to the $M_{\rm BH}/M_{\rm gal}$ relation.  This is consistent
with the observation that the majority of SMBH growth density (using \xray\
emissivity as a proxy) in the Universe at $z \approx 2$ can be attributed to
luminous AGN ($L_{\rm X} \simgt 10^{44}$~\lum; e.g., Hasinger \etal\ 2005); a
significant fraction of this population is expected to be fueled by mergers
(\hbox{$\approx$30--40\%}; e.g., Treister \etal\ 2012).  We note that none of
the luminous AGN among the C-COSMOS HAEs fell within the observational
footprint of the \hst\ WFC3 IR-imaged region of the the Cosmic Assembly
Near-infrared Deep Extragalactic Legacy Survey (CANDELS; Grogin \etal\ 2011;
Koekemoer \etal\ 2011), so it was not possible to examine directly the
rest-frame optical morphologies of the HAE AGN subpopulation.  We used the
\hst\ ACS optical (F814W) morphology catalogs from Tasca \etal\ (2009) from the
broader COSMOS field (Scoville \etal\ 2007) to compare the concentrations and
asymmetries of the rest-frame UV emission from the HAEs and their AGN
subpopulation.  These parameters were available for 270 C-COSMOS HAEs and all
10 AGN.  Compared with the C-COSMOS HAE population, the AGN subpopulation
morphology distribution is skewed towards higher optical-light concentrations
(K-S test reveals the populations differ at the 98.1\% confidence level).  The
AGN subpopulation asymmetry distribution is consistent with the broader HAE
population, with no evidence for more luminous AGN having larger asymmetries,
as might be expected from merging systems.  However, the observed \hst\ images
reveal that the rest-frame UV light from the most luminous AGN is dominated by
the QSO emission, skewing the light to higher concentrations and making merger
signatures elusive.  Observations of these sources with \hst\ WFC3 IR
(rest-frame optical) would improve our ability to address whether mergers play
a significant role in triggering driving SMBH growth in the HAE population.

\section{Summary}

We have conducted a $\approx$100~ks \chandra\ observation over the 2QZ~Clus
structure and present a \chandra\ point-source catalog.  Within the \chandra\
footprint, the 2QZ~Clus structure contains 21 HAEs and four QSOs at $z = 2.23$
(total of 22 unique $z = 2.23$ sources) and is on-average a factor of
$\approx$2 times overdense compared to field HAEs.    The
2QZ~Clus was initially selected by (1) identifying an overdensity of $z \approx
2.23$ QSOs and (2) performing H$\alpha$ narrow-band photometry to
preferentially detect additional star-forming galaxies and AGN at $z \approx
2.23$ (via HAE selection; see Matsuda \etal\ 2011).  Given its selection, the
2QZ~Clus HAEs contain a rich mixture of AGN and star-forming galaxies.  To put
into broader context the 2QZ~Clus HAEs, we compare their properties with those
of a larger sample of 210 $z = 2.23$ HAEs selected in the C-COSMOS field.  Our
findings are summarized below.

\begin{itemize}

\item We find seven of the 22 $z = 2.23$ sources are detected in the \xray\
band, including all four QSOs.  These sources have \hbox{2--10~keV}
luminosities of $\approx$(8--60)~$\times 10^{43}$~\lum\ and HB/SB count-rate
ratios indicative of unobscured QSOs.  Comparison with HAEs in \hbox{C-COSMOS}
reveal that the HAE AGN fraction in the 2QZ~Clus field is enhanced by a factor
of \hbox{$\approx$$3.5^{+3.8}_{-2.2}$} compared with the broader field (after excluding
pre-selected 2QZ~Clus QSOs).  Therefore, the 2QZ~Clus structure is a region in
which SMBH growth is in a particularly active phase.  

\item On average, X-ray selected AGN and QSOs occupy the high-flux end of the
H$\alpha$ flux distribution, suggesting that the H$\alpha$ line for these
bright HAEs is on average dominated by the QSO component.  For 2QZ~Clus the
mean H$\alpha$ flux is higher than any non-AGN HAE, indicating that the total
2QZ~Clus H$\alpha$ power output for HAEs is dominated by AGN/QSO activity.
However, the C-COSMOS HAEs have H$\alpha$ power dominated by star-forming
galaxies.

\item We find suggestive evidence that the C-COSMOS \hbox{$z = 2.23$} HAE AGN
fraction increases with increasing local HAE overdensity.  The 2QZ~Clus HAEs
reside in a factor of $\approx$2 HAE overdense environment.  After excluding
known optically-selected QSOs, the 2QZ~Clus AGN fraction is a factor of
$\approx$$4.1^{+5.0}_{-2.8}$ times higher than C-COSMOS HAEs in similarly
overdense environments.  Either the local density as measured by HAEs alone
does not provide an accurate characterization of a truly more overdense
2QZ~Clus environment, or the 2QZ~Clus HAE population is a statistical outlier
(at the $\approx$95.9\% level) in terms of its elevated AGN fraction.

\item We make use of \chandra\ and \herschel\ 250$\mu$m stacking of 2QZ~Clus
and C-COSMOS HAE samples to measure mean SMBH mass accretion rates and SFRs.
These stacking analyses reveal that the mean \xray\ emission from all HAE
samples is dominated by AGN, with 2QZ~Clus HAEs having QSO-like luminosities
($L_{\rm 2-10~keV} \approx$~[6--10]~$\times 10^{43}$~\lum) and C-COSMOS HAEs
having much lower Seyfert-like luminosities ($L_{\rm 2-10~keV}
\approx$~[3--6]~$\times 10^{42}$~\lum).  The mean SFRs range from
$\approx$60--110~\sfr\ for 2QZ~Clus HAEs depending on whether
optically-selected QSOs are removed or not.  The SFRs for C-COSMOS HAEs span
$\simlt$40~\sfr\ to $\approx$70~\sfr.  The inferred mean SMBH mass accretion
rate to SFR ratio ($\dot{M}_{\rm BH}$/SFR) for these samples indicate that the
2QZ~Clus SMBHs and galaxies appear to be growing at rates comparable to those
expected by the local $M_{\rm BH}/M_{\rm gal}$ ratio, while the mean
$\dot{M}_{\rm BH}$/SFR for C-COSMOS HAEs lie a factor of $\approx$3 times lower
than the local $M_{\rm BH}/M_{\rm gal}$ ratio.  We find that the $\dot{M}_{\rm
BH}$/SFR vs.~\xray\ luminosity for both the 2QZ~Clus and C-COSMOS HAE samples
appear to follow trends found for $z \approx$~1--2 star-forming galaxies and $z
\approx 2$ \xray\ selected AGN.  

\item We estimate that an episodic QSO phase with a duty cycle of
$\approx$2--8\% would allow C-COSMOS HAEs to emerge onto the $M_{\rm BH}/M_{\rm
gal}$ relation despite their lower population-averaged $\dot{M}_{\rm BH}$/SFR.
This estimate is consistent with the observed C-COSMOS HAE AGN fraction
($\approx$0.4--2.3\%) for $L_{\rm X} \simgt 10^{44}$~\lum\ sources.

\end{itemize}

Future observations of the 2QZ~Clus are needed to further evaluate (1)
what physical conditions have led to such an enhanced AGN fraction, (2) the
large-scale environment of the 2QZ~Clus structure, and (3) whether there is a
significant underlying overdensity of galaxies that are not HAEs (e.g., massive
passive galaxies and additional galaxies with low SFRs or heavily extinguished
H$\alpha$ emission).

\acknowledgements

We thank the anonymous referee for reviewing the manuscript and providing
helpful suggestions.  We gratefully acknowledge financial support from
\chandra\ X-ray Center grant GO2-13138A (B.D.L., A.B.L.), the Science and
Technology Facilities Council (STFC) (I.R.S.) and the Leverhulme Trust (I.R.S.,
J.R.M.).

%

%

\end{document}